\documentclass[8pt,a4paper]{article}
\usepackage[top=15truemm,bottom=15truemm,left=15truemm,right=15truemm]{geometry}
\usepackage{ascmac}
\usepackage{authblk}
\usepackage{bm}
\usepackage{braket}
\usepackage[
singlelinecheck=false,
font=sf,
]{caption}
\usepackage{color}
\usepackage[sharp]{easylist}
\usepackage{fancybox}
\usepackage{framed}
\usepackage{float}
\usepackage{graphicx}
\usepackage{longtable}
\usepackage{here}
\usepackage{hyperref}
\usepackage[version=3]{mhchem}
\usepackage{multicol}
\usepackage{multirow}
\usepackage{newtxtext}
\usepackage{multicol}
\usepackage{pdflscape}
\usepackage{pdfpages}
\usepackage{textcomp}
\usepackage{titlesec}
\usepackage[table,xcdraw]{xcolor}

\usepackage[
backend=biber,
style=phys,
sortcites=true,
url=false,
isbn=false,
date=year,
autocite=superscript,
natbib=true,
]{biblatex}
\titleformat*{\section}{\large\bfseries}
\titleformat*{\subsection}{\normalsize\bfseries}
\titleformat*{\subsubsection}{\normalsize\itshape}

\renewcommand{\figurename}{{Supplementary Figure}}
\renewcommand{\tablename}{{Supplementary Table}}

\makeatletter
\renewcommand{\maketitle}{
  \noindent
    {\@title} \\  
    {\@author} \\     
}
\makeatother

\newcommand{\reffig}[1]{\textbf{\figurename\ \ref{#1}}}
\newcommand{\reftab}[1]{\textbf{\tablename\ \ref{#1}}}

\makeatletter
\def\fnum@figure{\textbf{\figurename~\thefigure}}
\def\fnum@table{\textbf{\tablename~\thetable}}
\makeatother

\hypersetup{colorlinks=true,citecolor=black,linkcolor=black,urlcolor=black}

\begin{document}


\title{
{\normalsize 
  \hspace{-2mm}
  \textit{Supplementary Information for}
}\\
{\Large 
  \hspace{-1.65mm}
  \textbf{New metastable ice phases via supercooled water}\vspace{0.5cm}
}
}
\author[1,*]{Hiroki Kobayashi}
\author[1]{Kazuki Komatsu}
\author[2]{Kenji Mochizuki}
\author[1]{Hayate Ito}
\author[3]{Koichi Momma}
\author[4]{Shinichi Machida}
\author[5]{Takanori Hattori}
\author[6]{Kunio Hirata}
\author[6]{Yoshiaki Kawano}
\author[6]{Saori Maki-Yonekura}
\author[6,\dag]{Kiyofumi Takaba}
\author[6,7]{Koji Yonekura}
\author[2]{Qianli Xue}
\author[1]{Misaki Sato}
\author[1]{Hiroyuki Kagi}

\affil[1]{\footnotesize Geochemical Research Center, Graduate School of Science, the University of Tokyo, 7-3-1 Hongo, Bunkyo-ku, Tokyo, 113-0033, Japan}
\affil[2]{Department of Chemistry, Zhejiang University, Hangzhou 310058, People's Republic of China}
\affil[3]{National Museum of Nature and Science, 4-1-1 Amakubo, Tsukuba, Ibaraki, 305-0005, Japan}
\affil[4]{Neutron Science and Technology Center, Comprehensive Research Organization for Science and Society (CROSS), 162-1 Shirakata, Tokai-Mura, Naka, Ibaraki, 319-1106, Japan}
\affil[5]{J-PARC Center, Japan Atomic Energy Agency, 2-4 Shirakata, Tokai-Mura, Naka, Ibaraki, 319-1195, Japan}
\affil[6]{SPring-8 Center, RIKEN, 1-1-1 Koto, Sayo-Cho, Sayo-Gun, Hyogo, 679-5148, Japan}
\affil[7]{Institute of Multidisciplinary Research for Advanced Materials, Tohoku University, Aoba, Sendai, Miyagi 980-8577, Japan}




# New metastable ice phases via supercooled water

Hiroki Kobayashi[1,*], Kazuki Komatsu[1], Kenji Mochizuki[2], Hayate Ito[1], Koichi Momma[3], Shinichi Machida[4], Takanori Hattori[5], Kunio Hirata[6], Yoshiaki Kawano[6], Saori Maki-Yonekura[6], Kiyofumi Takaba[6,†],

Koji Yonekura[6,7], Qianli Xue[2], Misaki Sato[1], and Hiroyuki Kagi[1]

[1]Geochemical Research Center, Graduate School of Science, The University of Tokyo, 7-3-1, Hongo, Bunkyo-ku, Tokyo, 113-0033, Japan; [2]Department of Chemistry, Zhejiang University, Hangzhou 310028, People's Republic of China; [3]National Museum of Nature and Science, 4-1-1 Amakubo, Tsukuba, Ibaraki, 305-0005, Japan; [4]Neutron Science and Technology Center, Comprehensive Research Organization for Science and Society (CROSS), Tokai-Mura, Naka, Ibaraki, 319-1106, Japan; [5]J-PARC Center, Japan Atomic Energy Agency, Tokai-Mura, Naka, Ibaraki, 319-1195, Japan; [6]SPring-8 Center, RIKEN, 1-1-1, Koto, Sayo-Cho, Sayo-Gun, Hyogo, 679-5148, Japan; [7]Institute of Multidisciplinary Research for Advanced Materials, Tohoku University, Aoba, Sendai, Miyagi 980-8577, Japan

†Now at Institut für Anorganische Chemie, Fakultät für Chemie, Universität Wien, Währinger Straße 42, A-1090 Wien, Austria

*Corresponding author. E-mail: hiroki@eqchem.s.u-tokyo.ac.jp

## ABSTRACT

Water exhibits rich polymorphism, where more than 20 crystalline phases have been experimentally reported. Five of them are metastable and form at low temperatures by either heating amorphous ice or degassing clathrate hydrates[1–3]. However, such metastable phases rarely crystallise directly from liquid water, making it challenging to study metastable phase relations at relatively high temperatures. Here, we report that high-pressure metastable phases of ice, including two unknown phases named ices XXI and XXII, crystallise directly from liquid water in a deeply supercooled region around the homogeneous nucleation temperature. The key is to use emulsified water to stabilise supercooled water in laboratory timescales. Ices XXI and XXII are obtained by isothermal compression of emulsified water at 295 K and 250 K, respectively. Our powder x-ray and neutron diffraction analyses combined with molecular dynamics (MD) simulations revealed the surprisingly complex structures of these new phases with $Z = 152$ (ice XXI) and 304 (ice XXII). Ice XXI is topologically identical to 'ice T2' previously predicted by MD simulations, and our experimental structural model can be used as a benchmark for its structures in simulations, which depend on the force fields. On cooling, ice XXI transforms into an orientationally ordered counterpart named ice XXIII. Our results revealed the "hidden" structural complexity of water underlying the phase diagram, as implied by previous computational works. Further efforts at unveiling such metastable phase relations will bridge the large gaps between computational and experimental phase diagrams of water.

Water's molecular structure and hydrogen bonds are flexible. This leads to the enormous family of ice structures involving various metastable phases. The most accessible metastable form of ice is ice $I_{sd}$, also called stacking-disordered ice. It is a molecular-scale mixture of hexagonal and cubic stacking patterns, which is obtained by various methods such as freezing supercooled water near the homogeneous nucleation temperature ($T_H$) and heating amorphous ices at 1 atm[4–9]. Nevertheless, less is known about the crystallisation of metastable phases at high pressures. It is acknowledged in the community that high-pressure metastable phases are hardly prepared by simply freezing water. Metastable ice IV, for example, rarely crystallises from liquid water, and is therefore called a "will-o'-the-wisp", a tentative, ghostly form of ice[10–13]. Another high-pressure metastable phase, ice XII, was reported to form from liquid at 0.55 GPa and around 260 K, but required a cooling rate strictly controlled at 2.5 K/hour[14]. It has been more common and established to prepare ices IV and XII via ice $I_h$[1,15,16]. Moreover, numerous ice phases have been predicted in computer simulations but have not been identified experimentally[17–25], being leading candidates for new metastable phases. It is, therefore, of key importance to establish the crystallisation pathways from liquid water to dense metastable phases, for advancing experimental exploration of water's phase relations.

Based on Ostwald's step rule, we hypothesised that metastable ice phases were obtained from liquid water supercooled as deep as possible around $T_H$ (Figure 1a). To probe this, experiments must have been designed to achieve the deepest supercooling possible in bulk water. Emulsified water, where (sub-)micrometre-sized water droplets are dispersed in an organic oil matrix with surfactants (Figure 1b, Extended Data Figure 1), is a well-established sample widely used to reach $T_H$ since Pound et al. (1953) reported the homogeneous nucleation of ice I from supercooled emulsified water at 233 K, 1 atm[6,26–32]. We adopted the emulsification technique in our high-pressure experiments to study the crystallisation behaviour of supercooled water around $T_H$.

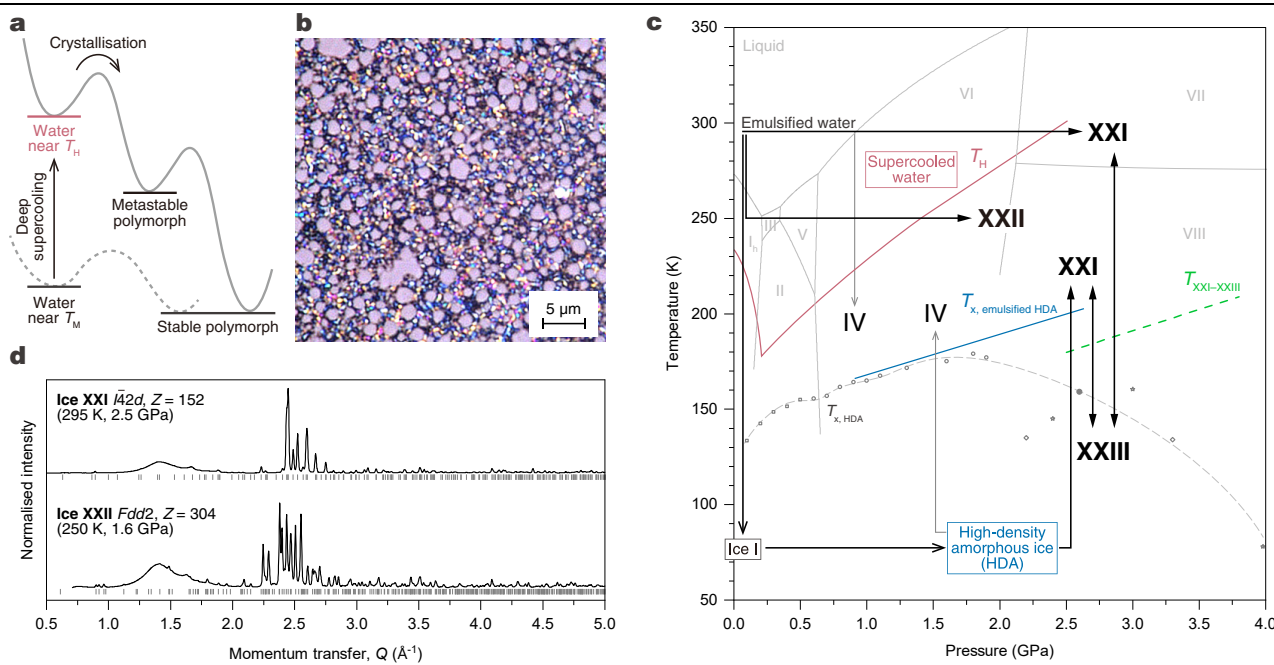

**Figure 1 | Summary of the crystallisation experiments.**
**a**, Schematic potential energy surface involving the crystallisation of a metastable ice polymorph from deeply supercooled water around the homogeneous nucleation temperature, $T_H$. **b**, Optical microscope image of an emulsified water sample (water:matrix = 7:10 volume ratio, see Methods). **c**, Pressure-temperature pathways from emulsified water to metastable ice phases shown on the phase diagram of water. Solid grey curves represent accepted phase boundaries between thermodynamically stable phases[33,34]. The crystallisation temperatures ($T_x$) of high-density amorphous ice (HDA) are shown by open squares (Seidl et al., 2013[35], 2015[36]), open circles (Stern et al., 2019[37]), open stars (Hemley et al., 1989[38]), open diamonds (Yoshimura et al., 2006[39], 2011[40]) and a filled grey circle (this study, Extended Data Figure 4d), with a dashed curve solely as a guide to the eye. **d**, In-situ powder x-ray diffraction patterns of ices XXI (top; 295 K, 2.5 GPa) and XXII (bottom; 250 K, 1.6 GPa). Tick marks represent the calculated Bragg-peak positions based on the determined unit cell. The broad feature centred at $Q = \sim 1.4$ Å$^{-1}$ is a characteristic halo peak of the organic matrix phase of emulsified water.

## Crystallisation paths

We begin with the overview of our crystallisation experiments using emulsified water, summarised in Figure 1c. When compressed at 295 K, emulsified water exhibited a substantial degree of overcompression, skipped stable ices VI and VII, and finally crystallised into a new phase named ice XXI at ~2.4 GPa (Figure 1c,d). When compressed at 250 K, at which the crystallisation pressure decreased to ~1.5 GPa, emulsified water crystallised into another new phase named ice XXII (Figure 1c,d). At lower temperatures (and pressures), emulsified water reproducibly crystallised into metastable ice IV at 0.7–1.1 GPa (Extended Data Figure 2). This is the first established method to obtain single-phase ice IV by freezing liquid water, which is surprising to us given its "will-o'-the-wisp" nature.

Both new phases form reproducibly and are stable in the order of days, enabling crystallographic characterisation by x-ray and neutron diffraction methods. They can also be recovered to 1 atm at liquid nitrogen temperature (Extended Data Figure 3).

Furthermore, ices IV and XXI can be obtained via alternative pathways, involving heating emulsified high-density amorphous ice (HDA). Given that heating HDA leads to the formation of metastable crystalline phases[1,37], it is of natural curiosity how different its crystallisation behaviour is in emulsified systems. It turned out that emulsified HDA crystallised by heating into ices IV and XXI below and above ~2 GPa, respectively (Extended Data Figure 4a). As with emulsified water, ice nucleation is suppressed in emulsified HDA; at 2.5 GPa, the crystallisation temperature ($T_X$) of emulsified HDA was ~200 K [into ice XXI], higher by ~40 K than $T_X$ of non-emulsified HDA [into ice VII] (Extended Data Figure 4c,d). We attribute the preferential crystallisation of ice XXI from overheated HDA to the difference in the crystallisation rates, analogous to the ice IV/XII formation at 0.81 GPa[1,41] (Extended Data Figure 4e).

## Structure of ice XXI

*In-situ* x-ray diffraction patterns of ice XXI were indexed by a large tetragonal unit cell of $a = $ ~19.7 Å and $c = $ ~7.3 Å. Structure determination by powder diffractometry (SDPD) was the only viable option for small crystals prepared from emulsified water, but it proved extraordinarily challenging due to the heavy overlap of Bragg reflections. To overcome such difficulties, we employed the maximum entropy method (MEM) for the Patterson function density to improve the deconvolution of heavily overlapped peaks and to estimate unobserved reflections in high-$Q$ regions, as demonstrated by Palatinus *et al.* (2007)[42]. Successive charge-flipping analysis revealed an oxygen-sublattice structure with $Z = 152$, which reproduced the experimental x-ray diffraction patterns well.

It is essential to examine how hydrogen bonds are arranged to construct this complex network structure. It was, however, impossible to determine the positions of more than 600 deuterium atoms based solely on experimental neutron diffractograms. Therefore, we performed molecular dynamics (MD) simulations to add hydrogen atoms to our model, starting from the experimentally determined oxygen positions. Simulations were performed using the TIP4P/Ice and TIP5P potentials[43,44]. These potentials overall suggested similar structures, where all water molecules formed a tetrahedral hydrogen-bonded network without destructive movements in oxygen atoms, but with specific differences in the network around the molecules labelled W6 (Figure 2a–d). TIP5P suggested that the network around W6 was highly disordered with several types of W6–W6 bonds, whilst TIP4P/Ice suggested a more ordered structure with fewer W6–W6 bonds. Interestingly, neutron diffraction data helped evaluate which computational suggestion was more plausible. The best Rietveld fit was achieved by an $I\overline{4}2d$ model with fewer W6–W6 bonds, as suggested by TIP4P/Ice (Figure 2e,f). We note that crystallographically, the $I\overline{4}2d$ space group can describe both structures suggested by TIP4P/Ice and TIP5P. However, even when the TIP5P suggestion was used as an initial model, the "wrong" deuterium sites move towards the "correct" sites (ultimately to be merged) and/or the occupancies of the "wrong" sites become nearly zero throughout Rietveld refinement, leading to more ordered structures similar to the TIP4P/Ice suggestion. This demonstrates the firm suitability of the TIP4P/Ice structure. The refined structure consists of a zig-zag chain of W6 molecules. It was remarkable that neutron diffraction experiments provided us with such

computational insight, where TIP4P/Ice worked better than TIP5P in terms of computationally reproducing the structure of ice XXI.

Importantly, ice XXI has almost the same structure as 'ice T2' predicted in MD simulations by Yagasaki *et al.* (2018)[21]. As Figure 2g shows, the proposed oxygen positions of ice T2 match the electron density map computed using our structural model of ice XXI. In simulations, ice T2 spontaneously nucleates at 260 K and 3.3 GPa and has a stability region between those of ices VI and VII in the phase diagrams of water. Ice XXI is the very first polymorph of ice crystallised in experiments at pressure-temperature conditions similar to where computational predictions were made beforehand.

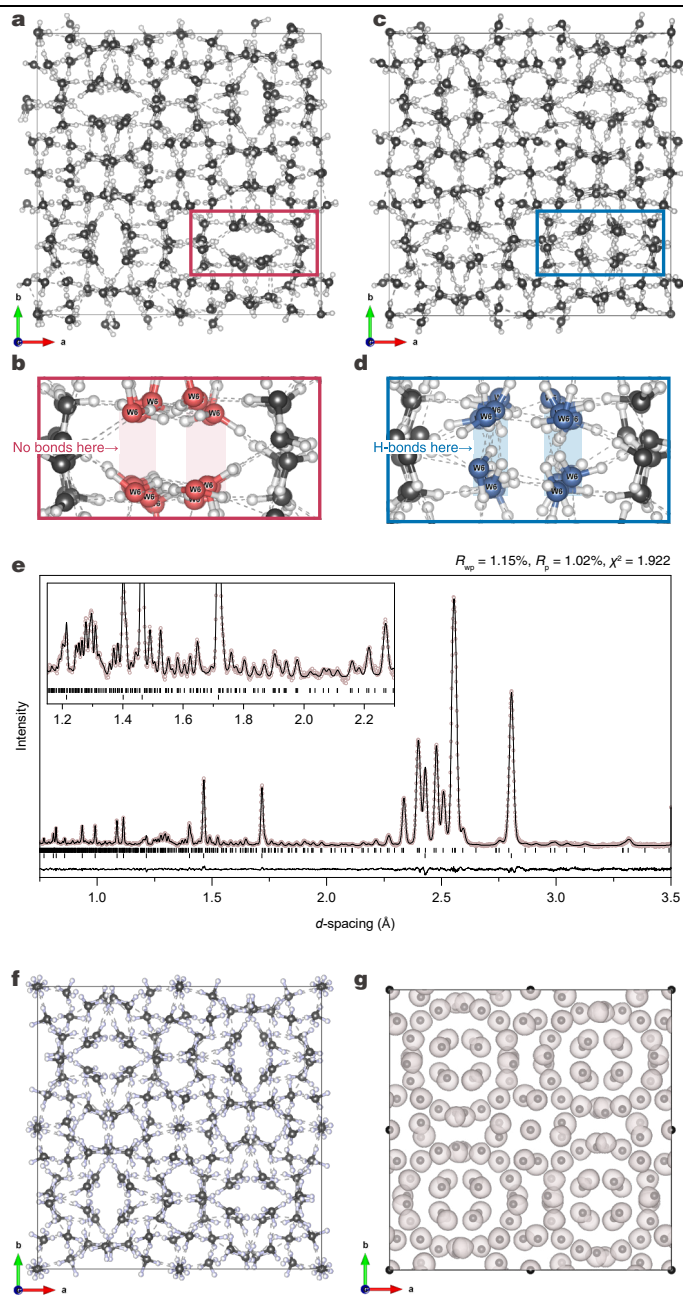

**Figure 2 | Ice XXI.**
**a–d**, Snapshots from MD simulations of ice XXI using the TIP4P/Ice (**a,b**) and TIP5P (**c,d**) potentials, starting from the experimental oxygen positions. Panels **b** and **d** show enlarged W6–W6 networks. **e**, Observed, calculated, and difference neutron diffraction profiles for ice XXI at 2.55 GPa, 204 K (background subtracted). The calculated fit (solid curve, $I\bar{4}2d$ model) is shown overlying the observed data (circles), and the residual is shown below. Tick marks represent reflection positions of ice XXI (upper) and the lead pressure marker (lower). **f**, Refined structure of ice XXI viewed along the *c*-axis. **g**, Model electron density of ice XXI (clouds, calculated from the model shown in **f**) and the oxygen positions of ice T2 proposed by MD simulations by Yagasaki *et al.* (2018)[21] (balls). It is noted that the hydrogen positions in ice T2 were not reported.

# Structure of ice XXII

Ice XXII have an orthorhombic cell of about $15 \times 16.5 \times 25.5$ Å$^3$, which is even larger than that of ice XXI. This marks the largest record of the unit cell among the ice polymorphs. We solved the structure of ice XXII by similar SDPD protocols, employing Pawley fitting to x-ray data, optimisation of the Patterson map by MEM, charge flipping, MD simulations, and Rietveld fitting for neutron diffraction patterns (Figure 3a,b). Ice XXII has a tremendously complex crystal structure with $Z$ = 304 (20 symmetry-inequivalent molecules) and the $Fdd2$ space group symmetry. There are channel structures along the $c$-axis (Figure 3b), which consist of molecular arrays connected at "intersections" in between.

As far as we know, this structure has never been computationally predicted due to the gigantic unit cell, although a part of the structure might have appeared in predicted structures. We checked its stability in MD simulations by introducing the liquid water as a coexisting phase. Crystal growth of ice XXII was observed at 270 K, 1.65 GPa (Figure 3c–e), which certainly supported the stability of this structure.

We point out that this is not the first appearance of ice XXII in literature. As shown in Extended Data Figure 5a–b, our x-ray diffraction measurements revealed that ice XXII melted at the melting points of 'PNP-XIV' (PNP: possible new phase) reported based on the melting-curve measurements by Mishima and Stanley (1998)[32]. Since no known ice phases melt nearby, ice XXII and PNP-XIV are highly likely to be identical. Our MD simulations using the experimental structure model of ice XXII accurately reproduced the $p$-$T$ coordinates of the melting curve of PNP-XIV relative to that of stable ice VI (Extended Data Figure 5c), which also supported the direct relevance between ice XXII and PNP-XIV. According to Mishima and Stanley (1998), this phase is also obtained by 'heating emulsified HDA above $T_X$ around 1.5 GPa', but we have never observed this crystallisation pathway. As with the crystallisation of ice IV from non-emulsified HDA[1], there may be a strict pressure and/or heating-rate requirement to crystallise ice XXII from emulsified HDA, which would be a subject of future studies.

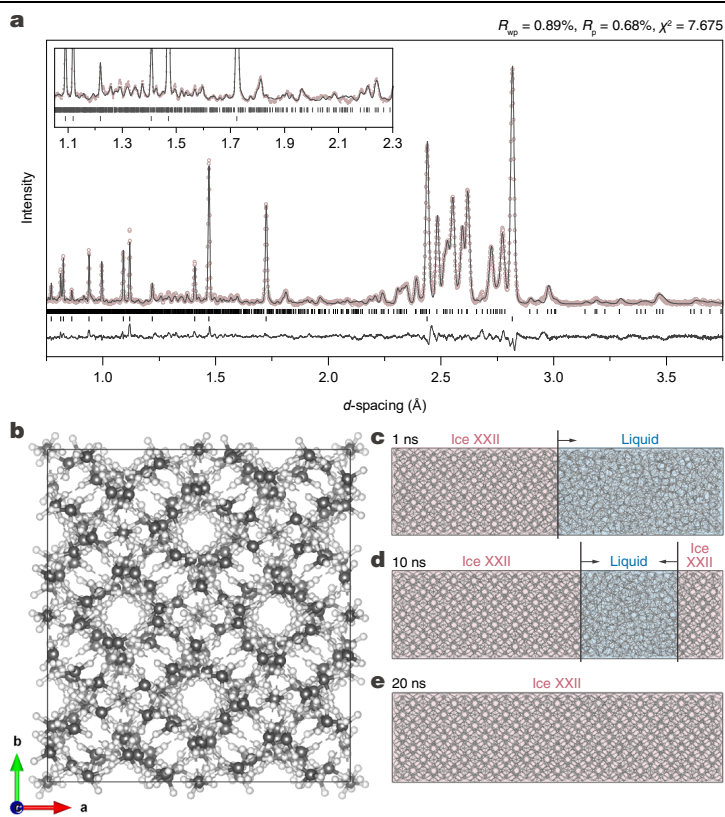

**Figure 3 | Ice XXII.**
**a**, Observed, calculated, and difference neutron diffraction profiles for ice XXII at 1.48 GPa, 93 K (background subtracted). The calculated fit (solid curve) is shown overlying the observed data (circles), and the residual is shown below. Tick marks represent reflection positions of ice XXII (upper) and lead (lower). **b**, Refined structure of ice XXII viewed along the $c$-axis. **c–e**, Snapshots during crystal growth of ice XXII in MD simulations at 270 K, 1.65 GPa, where the (100) plane was exposed to the liquid phase.

# Orientational ordering: ice XXIII

Ices XXI and XXII turned out to be hydrogen-disordered phases, in which each crystallographic oxygen site is surrounded by four hydrogen (deuterium) sites with an occupancy of 50% due to orientational disorder. Interest arises in how the structures of these phases change at low temperatures because orientational ordering is a common phenomenon in ice to achieve the lowest-energy hydrogen-bond configuration in a given oxygen sublattice at low temperature. We cooled ice XXI down to 93 K at a rate of ~0.2 K/min under pressure and tracked its structural changes by neutron diffraction. As shown in Figure 4 and Extended Data Figure 6, the lattice parameters showed abrupt changes associated with densification, indicating a first-order phase transition. The phase boundary has a positive slope in the $p$-$T$ diagram (Figure 4e), consistent with the volume reduction. We name this low-temperature phase ice XXIII. Discontinuous changes in diffraction intensities were observed (Figure 4f). Symmetry reduction is evident from neutron diffractograms because the positions of some new peaks were incompatible with the body-centred tetragonal lattice of ice XXI. $P2_12_12_1$ is the highest-symmetry subgroup of $I\bar{4}2d$ that gives a good Rietveld fit to the neutron diffractogram of ice XXIII (Figure 4g). All deuterium-site occupancies were refined since $P2_12_12_1$ permits a fully ordered structure. Rietveld analysis suggested that ice XXIII is only partially ordered, in which the occupancies of ten deuterium sites converged to either 0 or 1, but the deuterium atoms at the other sites remain less ordered (Extended Data Table 3).

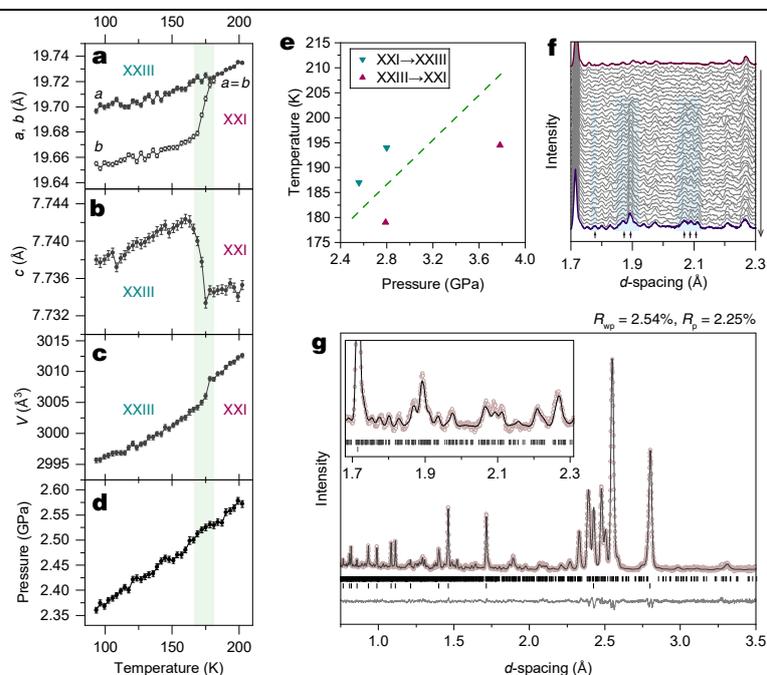

**Figure 4 | Hydrogen-ordering phase transition from ice XXI to XXIII.**
**a–d**, Lattice parameters ($a$, $b$ and $c$), unit-cell volume, and pressure during cooling at ~2.5 GPa. **e**, Onset points of the $D_2O$ ice XXI→XXIII and XXIII→XXI transitions estimated from the lattice parameters. **f**, Emergence of characteristic neutron diffraction peaks for ice XXIII. Patterns, except for the highest and lowest temperatures, were smoothed for better visibility. **g**, Observed, calculated, and difference neutron diffraction profiles for partially ordered ice XXIII at 93 K, 2.4 GPa (background subtracted). The calculated fit (solid curve) is shown overlying the observed data (circles), and the residual is shown below. Tick marks represent reflection positions of ice XXIII (upper, $P2_12_12_1$ model) and lead (lower).

In contrast, there were no clear indications of a phase transition in ice XXII during slow cooling (Extended Data Figure 7). To access the ordered counterpart of ice XXII, acid or base doping techniques may be effective, as demonstrated in ices $I_h$, IV, V, VI, and XII[45–50]. Doping may also be effective in obtaining more ordered ice XXIII, which would be essential to investigate its thermodynamic stability by, for example, performing theoretical calculations.

## Conclusions and outlook

Our experiments using emulsified water combined with MD simulations lead to four essential conclusions. First, in a wide pressure range between 0.7 and 2.4 GPa, deeply supercooled water reproducibly crystallises at around $T_H$ into metastable high-pressure phases of ice, including two new phases named ices XXI and XXII. Second, our structure analyses employing Le Bail/Pawley fitting, MEM, charge flipping, MD simulations and Rietveld refinement showed that ices XXI and XXII have very complex crystal structures with large unit cells, which proved previously unknown structural diversity of ice. Third, computationally predicted ice structures of ice are not necessarily "fictitious polymorphs" as ice XXI corresponds to ice T2 predicted by MD simulations under a similar condition previously. Since hydrogen-bond configurations in ice XXI are potential-dependent in simulations, our structure model is a useful constraint on choosing potentials. Fourth, the hydrogen disorder–order transition is experimentally observable without dopants for the ice XXI–XXIII pair, although ice XXIII achieves only partial ordering. All disorder–order transitions that do not require doping occur above 170 K, which may be a general criterion temperature for hydrogen-bond rearrangements in a fixed oxygen sublattice.

The discovery of the "computational" ice T2 stimulated us to search for other theoretically predicted ice structures, especially at higher temperatures[18,20,22,23,25]. To this end, we tested isothermal compression of emulsified water at 350 K, but it resulted in the stable phase, ice VII. This illustrates an expected fact that access to metastable phases becomes increasingly difficult as the temperature increases. Other methods, such as shock or dynamic compression and rapid quenching, would be powerful tools to further investigate metastable phase relations at higher temperatures.

This study demonstrated the general potential of supercooled liquids in obtaining metastable solids. We suggest that the use of small droplets, as well as other methods listed above and their combinations, has large applicability in broad materials science fields. These could be a clue to new materials in general, or even to unusual reactions in solutions facilitated thanks to high free energy in supercooled liquids. Such approaches would help with synthesising extraordinary materials, establishing polymorph selection methodologies, and bridging gaps between experiments and calculations.

## Methods

**Preparation of emulsified water.**

Emulsified water samples were prepared with matrix:water volume ratios of 10:7 or 10:2. Organic matrix (1:1 mixture of methylcyclohexane and methylcyclopentane) and powdered sorbitan tristearate (4.6 wt% of the matrix [the matrix:water = 10:7 sample] and 2.0–3.5 wt% of the matrix [the matrix:water = 10:2 sample]) were well mixed to obtain white suspensions (Extended Data Figure 1a,b). Thereafter, the prepared suspension and water (matrix:water = 10:7 or 10:2 volume ratio, see below) were mixed and homogenised under ice-cooled conditions by using an ultrasonic homogeniser (TAITEC, VP-050N; Extended Data Figure 1c,d). Ultrasonic waves were irradiated for 15–30 sec at the 35% output until the sample appeared homogeneous, followed by a second irradiation for 4 min at the 90% output power. Water particles in emulsified water had diameters of about 0.5–5 μm in the 10:7 sample and about 0.2–2 μm in the 10:2 sample (Figure 1b, Extended Data Figure 1e–h). Deuterated liquid samples were used for neutron diffraction.

**Crystallisation protocols of ices XXI and XXII.**

*Ice XXI from liquid water*

Emulsified water samples with a matrix:$H_2O$ volume ratio of 10:2 were prepared for this experiment. Emulsified water was compressed at room temperature up to ~2.4 GPa, or at 270 K up to ~2.0 GPa to yield ice XXI. With a larger amount of aqueous phase in emulsified water samples (matrix:water = 10:7 volume ratio), it is possible to obtain ice XXI through the same pathway, but ice-VII contamination is unavoidable.

*Ice XXI from HDA*

Emulsified water samples with a matrix:$H_2O$ volume ratio of 10:7 were prepared for this pathway. The emulsified liquid water was cooled at ~0 GPa to obtain ice $I_{sd}$. Thereafter, the obtained ice $I_{sd}$ was compressed to 2.5 GPa isothermally at temperatures below 135 K to obtain HDA[51]. HDA was heated up to 210 K at ~2.5 GPa. The heating rate was 1 K/min in *in-situ* diffraction experiments, and ~7 K/min in the piston-cylinder experiments for *ex-situ* measurements.

*Ice XXII from liquid water*

Emulsified water samples with a matrix:$H_2O$ ratio of 10:2 or 10:7 were used for this pathway. With the 10:7 sample, the ice-VI contamination was unavoidable, but ice VI was observed even with the 10:2 sample in some experiments. In angle-dispersive x-ray diffraction experiments using two-dimensional detectors, ice VI produced spots whilst ice XXII produced smooth Debye-Scherrer rings, and therefore the effects of contaminant Bragg peaks from ice VI could be minimised by proper masking before generating one-dimensional profiles. Emulsified water was cooled down to 250 K and compressed up to ~1.6 GPa there to prepare ice XXII. It is noted that emulsified water can be cooled at any pressure below 1.6 GPa because it never crystallises below 1.6 GPa and above 250 K. For example, ice XXII can also be obtained by compressing emulsified liquid up to ~1.6 GPa followed by cooling there.

***In-situ* diffraction measurements.**

High-pressure synchrotron x-ray diffraction experiments were carried out at BL-18C in Photon Factory of High-Energy Accelerator Research Organization (KEK) using diamond-anvil cells (DACs). The DACs we used consisted of a pair of Boheler-Almax-design diamond anvils (culet diameter: 600 μm or 800 μm) and a gasket (CuBe gasket with a thickness of 200 μm and a hole diameter of 300 μm, or stainless gasket with a thickness of 100 μm and a hole diameter of 300 μm) without pre-indentation. We placed DAC-loaded samples in a 4 K cryostat (Iwatani Co., MiniStat) where the temperature was controlled using a LakeShore 335 Temperature Controller. Sample pressure was estimated by the ruby-fluorescence method[52,53]. Fluorescence spectra were collected using a spectrometer (Princeton Instruments, Acton SpectraPro SP-2750) equipped with an EMCCD camera (ProEM 1600x200) whose pixel-wavelength relation was calibrated using emission lines

of neon. The sample was irradiated through an anvil with incident x-rays monochromatised at 20 keV using a Si (111) monochromator, and diffracted X-rays were detected either by a CMOS flat-panel detector (Rad-icon 2022) or by an imaging plate (Fujifilm, BAS-MS 2025). 2D x-ray images were converted to 1D profiles using *IPAnalyzer* software[54].

Some x-ray experiments were conducted at the laboratory at the University of Tokyo using a Rigaku R-AXIS IV++ diffractometer with Mo *Kα* radiation generated by a Rigaku MicroMax-007. DACs were equipped as described above and placed in a purpose-made cryostat (Japan Thermal Engineering Co.) using liquid nitrogen cryogen.

*In-situ* neutron diffraction experiments were performed at the high-pressure neutron diffractometer *PLANET*[55] in the Materials and Life Science Experimental Facility of J-PARC using a temperature-variable hydraulic press called *Mito system*[56]. Samples were encapsulated in a pair of gaskets made of null-scattering TiZr alloy and sandwiched by anvils made of tungsten carbide (for ice XXI/XXIII) or alumina-toughened zirconia (for ice XXII). A lead wire was loaded for pressure determination using the equation of state[57]. The sample was irradiated with white neutron pulses of 25 Hz through an anvil, and scattered neutrons were detected by $^3$He position-sensitive detectors on two 90° banks. Diffraction intensities were corrected using intensities of vanadium-filled and empty cells and normalised by the number of incident neutrons.

Information about each experiment is available in Supplementary Information.

**Quench recovery and *ex-situ* x-ray diffraction.**

For *ex-situ* x-ray characterisation, samples were prepared using piston-cylinder high-pressure cells with a bore diameter of 4 or 5 mm following the paths described above. After crystallisation, samples were cryo-quenched to 77 K and recovered to 1 atm under liquid nitrogen. The recovered samples were stored in liquid nitrogen and cryo-transported for *ex-situ* x-ray diffraction measurements. *Ex-situ* x-ray diffraction patterns were collected at BL32XU in SPring-8. Pieces of recovered samples were put on a Kapton holder and mounted to the diffractometer with dry nitrogen gas cryostream at 90 K. Monochromatised x-rays at 20 keV were used, and diffraction patterns were collected using a DECTRIS EIGER X 9M silicon diode detector. Collected data were analysed in similar procedures to *in-situ* experiments at BL-18C.

**Diffraction data analysis.**

*CONOGRAPH* software[58,59] was used for indexing and constraining space groups. The Le Bail and Pawley fitting was performed using *GSAS/EXPGUI*[60,61] and *Z-Rietveld*[62,63] software, respectively. The *ERIS* software was used for the optimisation of Patterson maps by the maximum entropy method (so-called maximum entropy Patterson (MEP) method). Rietveld analysis was performed using *GSAS/EXPGUI*[60,61]. *VESTA* was used for visualisation and generating low-symmetry structures[64]. The Bilbao Crystallographic Server, especially the 'MAXSUB' and 'SUBGROUPGRAPH' functions, was used to consider the group-subgroup relations[65–67]. Detailed descriptions can be found in the Supplementary Information.

**Molecular dynamics simulations.**

Molecular dynamics (MD) simulations were carried out in the *GROMACS* 2021.7 package[68]. Water molecules were represented by the TIP4P/Ice model[43] or the TIP5P model[44]. The equations of motion were integrated with the leapfrog algorithm using a time step of 2.0 fs. The non-bonded interactions were truncated at 0.90 nm, except for some calculations performed with 0.70 nm, see Supplementary Information. Long-range dispersion corrections were applied for energy and pressure. Long-range Coulombic interactions were evaluated with the particle-mesh Ewald (PME) algorithm. Periodic boundary conditions were applied in all directions.

Temperature was controlled by the stochastic velocity rescaling thermostat (called 'V-rescale' in *GROMACS*) with a damping constant of 0.1 ps. Pressure was controlled in three different ways. The stochastic cell rescaling barostat (called 'C-rescale' in *GROMACS*) with a damping constant of 1.0 ps was used for isotropic pressure control for liquid and a semi-isotropic pressure control for solid-liquid coexistence. For solids, to achieve an anisotropic pressure control, an equilibration

with the Berendsen barostat was followed by a production run with the Parrinello-Rahman barostat, both with a damping constant of 1.0 ps.

**Cryogenic transmission electron microscopic imaging.**

The size of the water droplets in the emulsified water with the matrix:water volume ratio of 10:2, which were hardly seen under optical microscopes, was checked using cryogenic electron microscopic imaging. Emulsified ice samples were crushed under liquid nitrogen and loaded onto glassy carbon films (QUANTIFOIL R1.2/1.3, 100 holey carbon films with Cu 200 mesh). Samples were quickly transferred into an electron microscope (JEOL CRYO ARM 300), maintained at 98 K and operated at 300kV, for observation.

# Crystallographic data availability

Until the peer-reviewed publication, the crystallographic information files for ices XXI, XXII, and XXIII are available at https://github.com/khsacc/New_ice_structures_arXiv.

# Acknowledgements

# Author contributions

# Extended Data

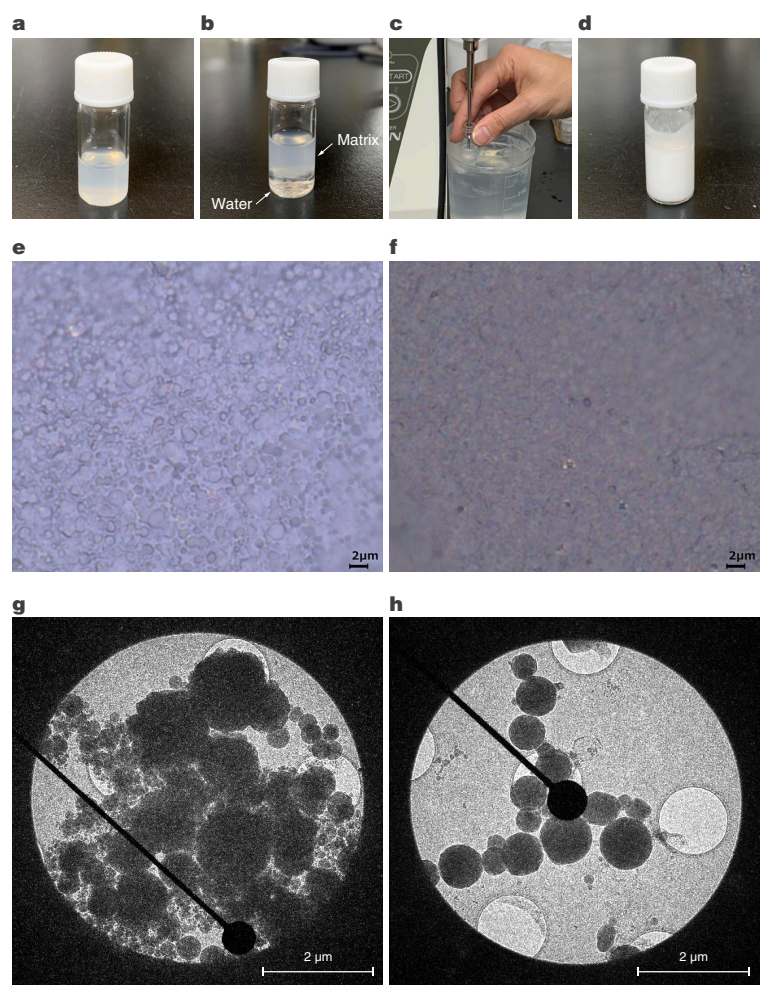

**Extended Data Figure 1 | Preparation and characterisation of emulsified water samples used in this study.**
**a-d**, Preparation protocol of emulsified water samples: (a) Surfactant powder (sorbitan tristearate) is well mixed with the organic matrix phase (1:1 mixture of methylcyclohexane and methylcyclopentane) by ultrasonic irradiation; (b) Water is added to the obtained white suspension; (c) Emulsification was performed by an ultrasonic homogeniser under an ice-cooled condition; (d) Prepared water-in-oil emulsion sample. **e–f**, Optical microscope images of emulsified water with the matrix:water volume ratios of (e) 10:7 and (f) 10:2. **g–h**, Cryogenic transmission electron microscopic images of frozen emulsified water (phase: ice $I_{sd}$) with the matrix:water = 10:2 volume ratio. The matrix phase evaporated in a vacuum condition in the electron microscope. Images were acquired in defocused diffraction mode with a beam stopper (stick and circle). The carbon hole diameter is ∼ 1 μm.

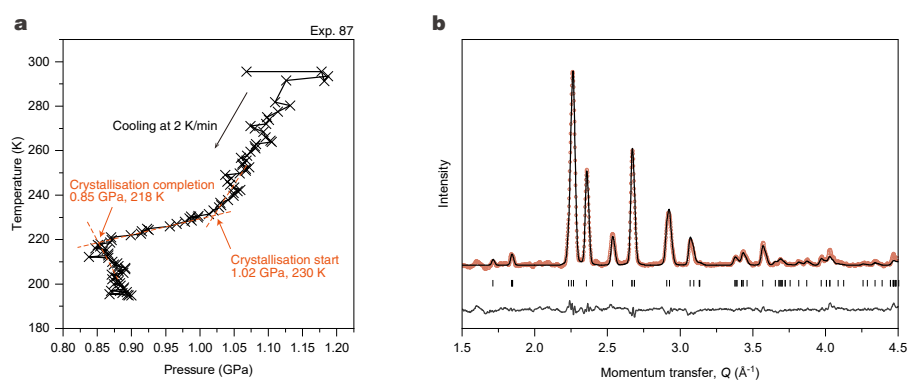

**Extended Data Figure 2 | Representative crystallisation experiment of ice IV from supercooled water.**
We only present representative data here and will report detailed results about ice IV elsewhere. This experiment was performed using a diamond anvil cell and a laboratory-source diffractometer (Rigaku R-AXIS IV++, Mo $K\alpha$ radiation). **a,** Pressure-temperature path, where pressure was estimated by the ruby fluorescence method[52,53]. Estimates of the crystallisation onset and completion points are shown aside. **b,** Observed, calculated, and difference x-ray diffraction profiles for ice IV crystallised directly from supercooled water (background subtracted). This diffraction pattern was measured at 194.5 K, 0.90 GPa. The curve through the circles (observed intensities) is a Rietveld fit. The difference profile is plotted below. Tick marks indicate the peak positions for ice IV.

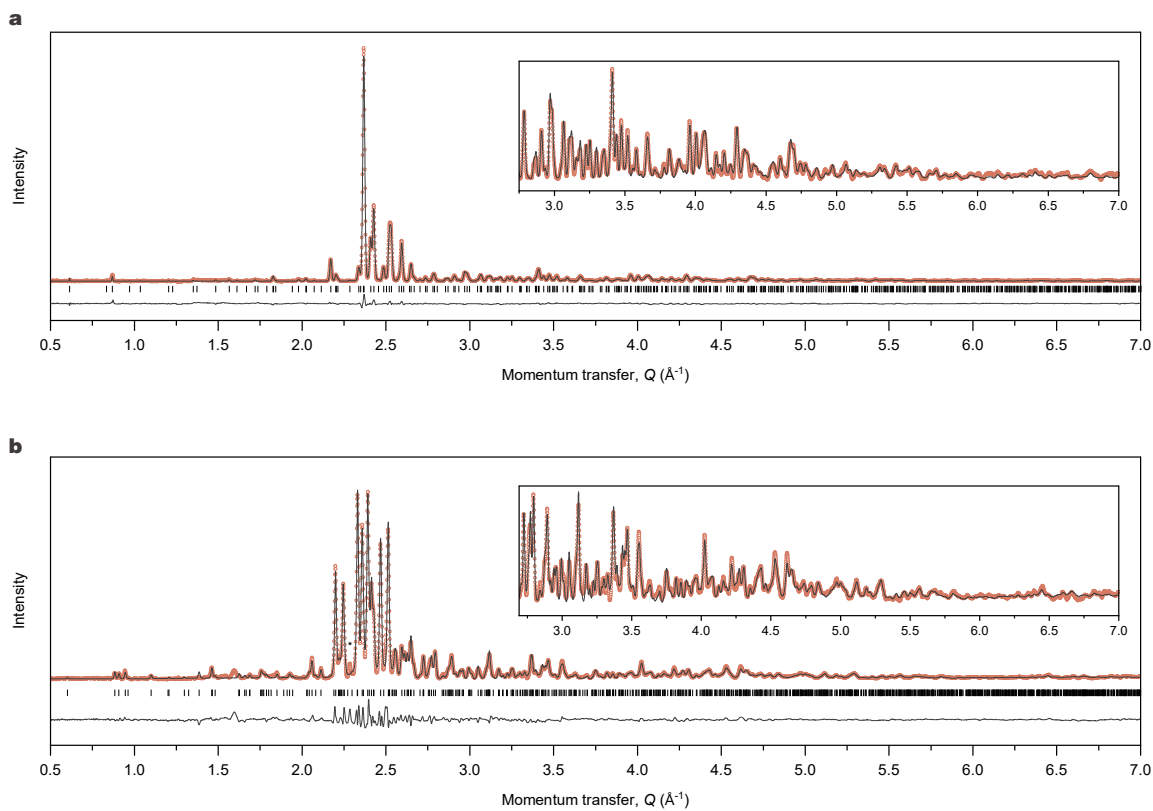

**Extended Data Figure 3 | Observed, calculated, and difference x-ray diffraction profiles for recovered ices (a) XXI (XXIII) and (b) XXII at 1 atm, 90 K.**
Data were collected at BL32XU. The background was subtracted in the plots. The curve through the circles (observed intensities) is a Rietveld fit starting from the structural model obtained by charge flipping calculations. The difference profile is plotted below. Tick marks indicate peak positions for the new phases. An unfitted peak indicated by an asterisk in panel **b** originates from contaminant ice VI.

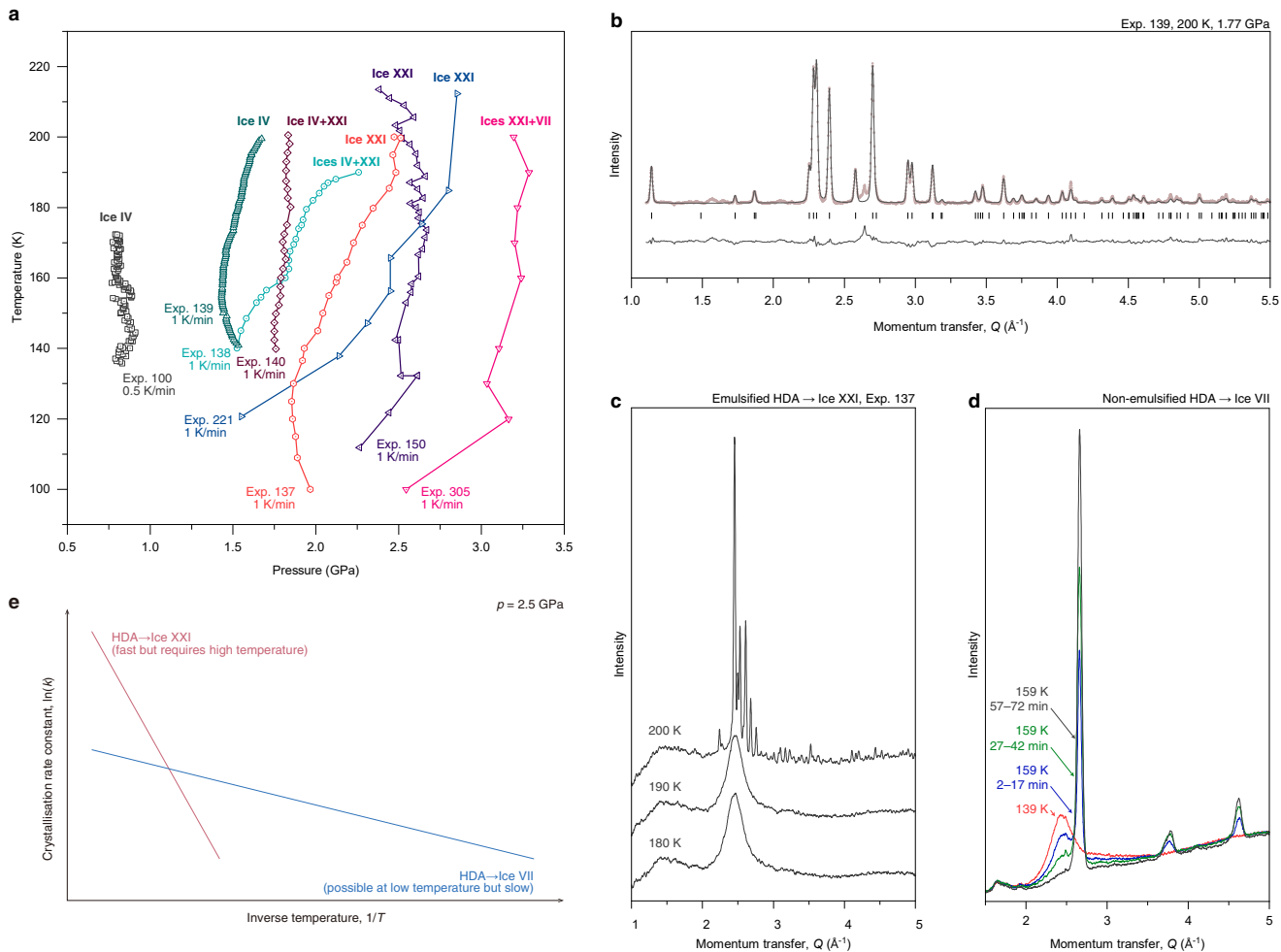

**Extended Data Figure 4 | Crystallisation of emulsified and non-emulsified HDA.**

**a**, Pressure-temperature paths we have tested to study the heat-induced crystallisation of emulsified HDA. Crystallisation products were identified by x-ray or neutron diffraction. **b**, Representative x-ray diffraction profile of emulsified ice IV crystallised from emulsified HDA at unusually high pressures (Exp. 139, ~1.8 GPa; synchrotron x-rays at 20 keV, BL-18C). The curve through the circles (observed intensities) is a Rietveld fit, and the difference profile is plotted below. Tick marks indicate the peak positions for ice IV. The unfitted peak centred at $Q = 2.65$ Å$^{-1}$ originated from ruby spheres loaded for pressure determination. **c**, Temperature-induced crystallisation of emulsified HDA at ~2.5 GPa tracked by x-ray diffraction (synchrotron x-rays at 20 keV, BL-18C). **d**, Temperature-induced crystallisation of non-emulsified HDA (uHDA) at ~2.5 GPa tracked by x-ray diffraction (Mo $K\alpha$ radiation, Rigaku R-AXIS IV$^{++}$). For the 159 K data, the time after reaching 159 K is shown aside; unlike the ice XXI crystallisation in emulsified systems, the crystallisation of ice VII from normal HDA took a relatively long time. **e**, Schematic illustration of the crystallisation rate constants at 2.5 GPa, constructed referring to a similar example of non-emulsified ices IV and XII at 0.81 GPa [Figure 4 in Ref. [1] by Salzmann *et al.* (2002) and Figure 10 of Ref. [41] by Tonauer *et al.* (2023)]. Quantitative analysis of the crystallisation kinetics of ices VII and XXI from emulsified HDA was not feasible in this study; however, it is evident that the crystallisation of ice XXI at ~200 K (**c**) is significantly faster than that of ice VII at 159 K (**d**). The coexistence of HDA and ice XXI has not been observed in our x-ray diffraction measurements of 1 min, while HDA and ice VII coexisted for about an hour.

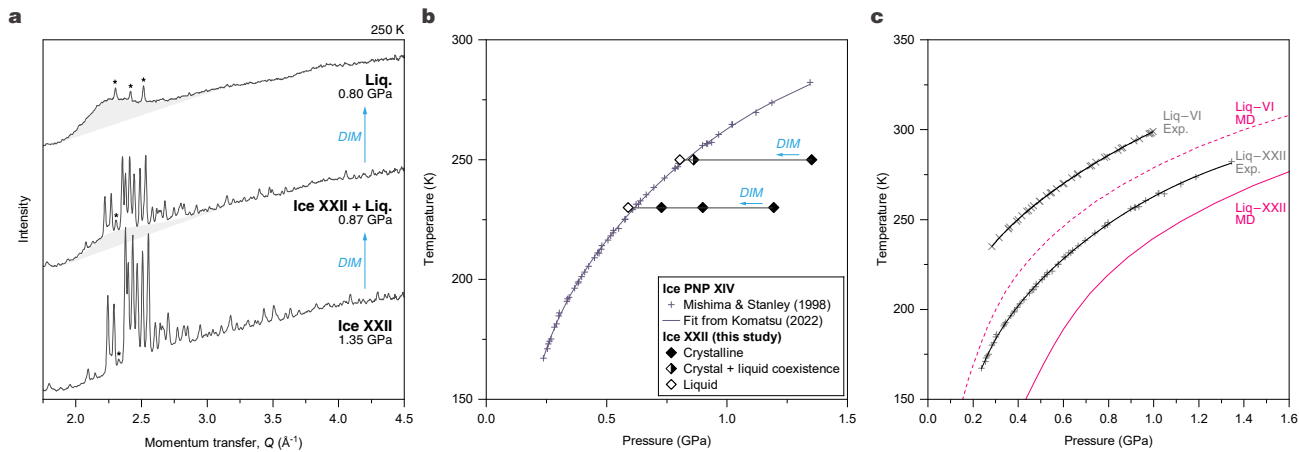

**Extended Data Figure 5 | Experimental and computational melting behaviour of ice XXII.**
**a**, X-ray diffraction patterns along isothermal decompression at 250 K, tracking decompression-induced melting (DIM) of ice XXII. Bragg peaks of contaminant ice VI are indicated by asterisks. **b**, DIM points compared between PNP-XIV (Mishima and Stanley, 1998; thermal measurements[32]) and ice XXII (this study; x-ray diffraction), showing the relevance of ice XXII to PNP-XIV. **c**, Melting curves of ices VI and XXII calculated by MD simulations. Experimental melting points of emulsified ice VI and PNP XIV (=XXII) reported by Mishima and Stanley (1998)[32] are also shown, along with the fitted curves by Komatsu (2022)[34]. Refer to the Supplementary Information for the calculation methods.

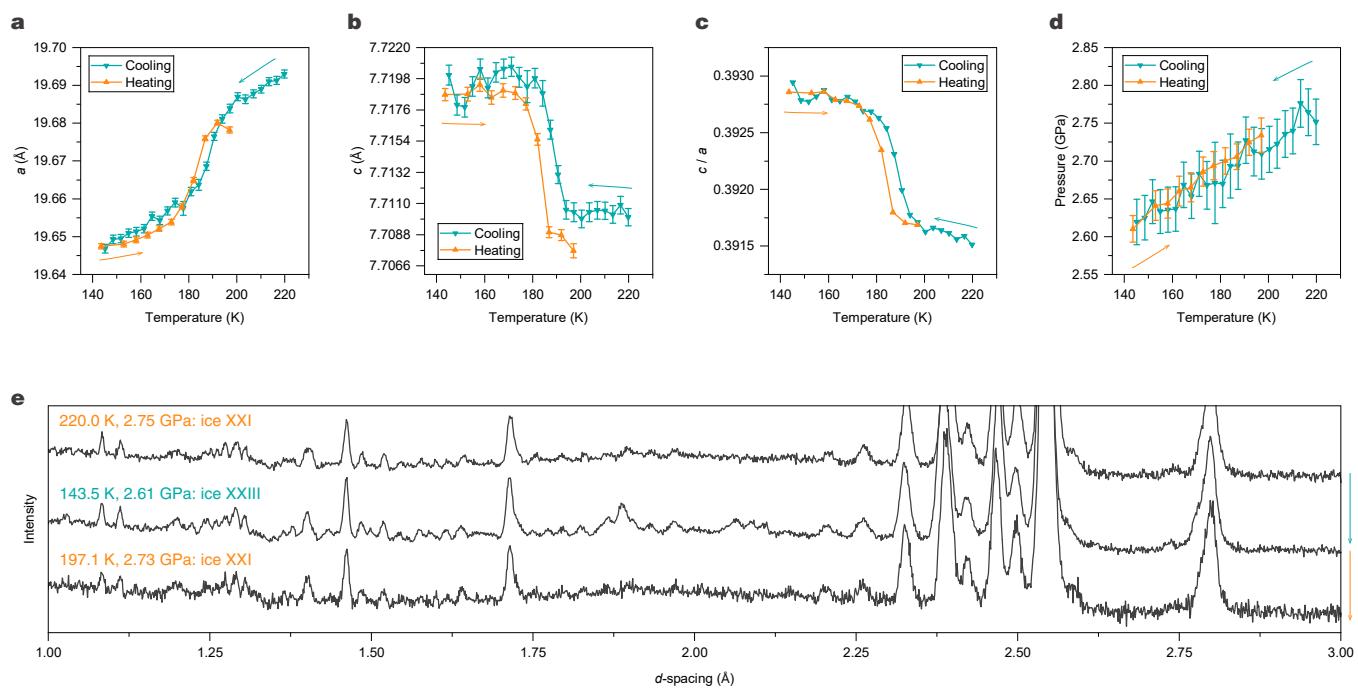

**Extended Data Figure 6 | Reversible ice XXI–XXIII transition at ~2.7 GPa.**
**a–d**, Lattice parameters, their ratio ($c/a$), and pressure estimated by the equation of state of lead during cooling and heating. **e**, Neutron diffraction profiles of as-crystallised ice XXI (top), ice XXIII obtained by cooling (middle), and ice XXI obtained by reheating (bottom).

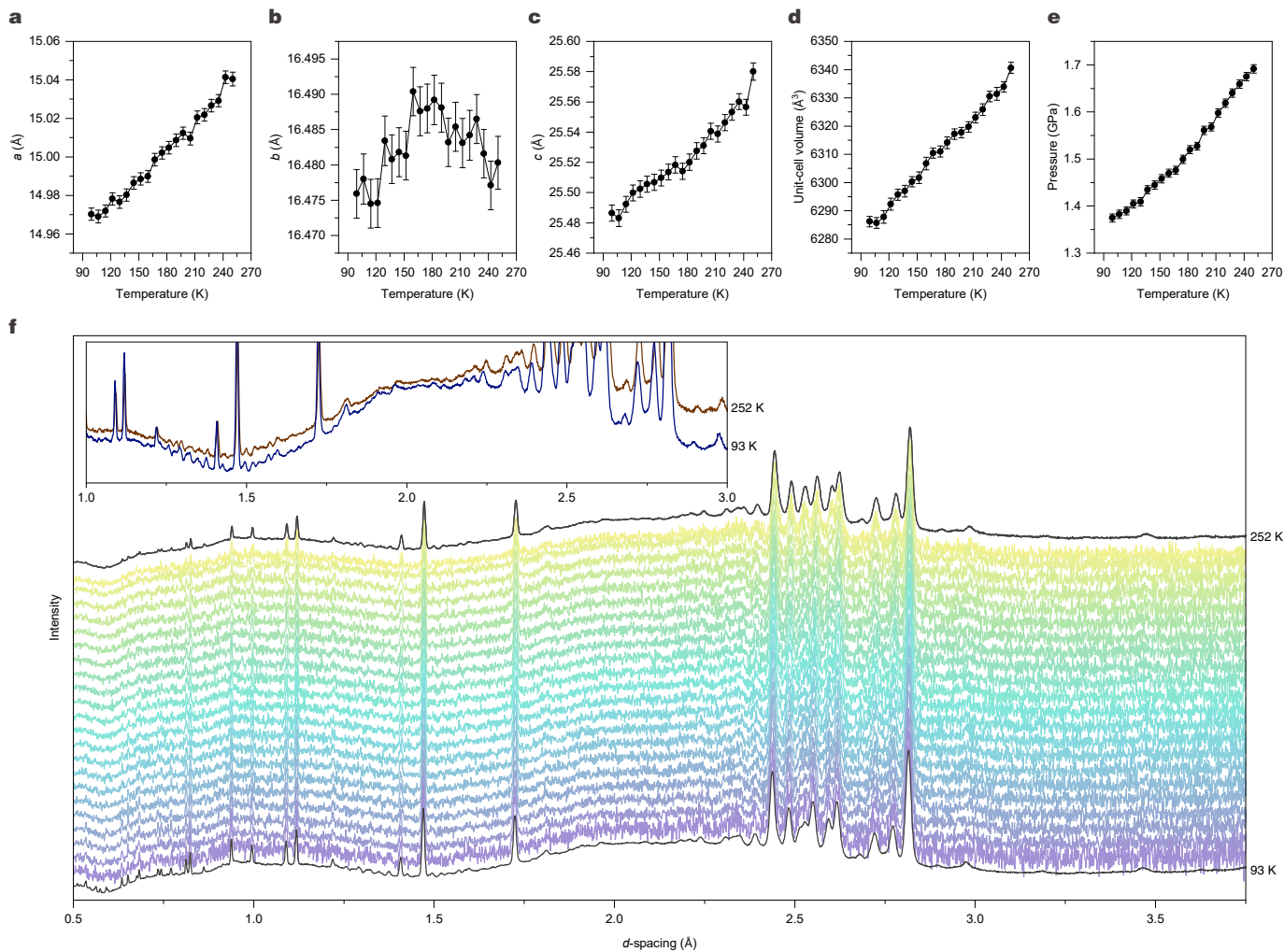

**Extended Data Figure 7 | Absence of a clear phase transition of ice XXII during cooling.**
**a–e**, Lattice parameters, unit-cell volume, and pressure estimated by the equation of state of lead during cooling. **f**, Neutron diffraction profiles during cooling.

**Extended Data Table 1 | Fractional coordinates, site occupancies, and isotropic atomic displacement parameters ($U_{iso}$) for ice XXI at 2.55 GPa, 204 K.**

| Site | $x$ | $y$ | $z$ | Occupancy | $U_{iso}$ |
|------|-----|-----|-----|-----------|-----------|
| O10 | 0 | 0 | 0.2363(4) | 1 | 0.0131(6) |
| O20 | 0.39317(20) | 0.11450(17) | 0.9966(5) | 1 | 0.0131(6) |
| O30 | 0.17930(17) | 0.08882(18) | 0.3558(6) | 1 | 0.0161(6) |
| O31 | 0.06963(18) | 0.17906(20) | 0.1289(5) | 1 | 0.0161(6) |
| O40 | 0.26216(19) | 0.07899(18) | 0.0225(5) | 1 | 0.0161(6) |
| O41 | 0.04165(18) | 0.26788(19) | 0.4505(5) | 1 | 0.0161(6) |
| O50 | 0.11523(21) | 0.03574(27) | 0.0632(5) | 1 | 0.0131(6) |
| O51 | 0.02136(19) | 0.10336(24) | 0.4793(4) | 1 | 0.0131(6) |
| O60 | 0.20322(20) | 0.17661(17) | 0.0453(6) | 1 | 0.0241(14) |
| O61 | 0.32616(23) | 0.20785(17) | 0.2940(5) | 1 | 0.0241(14) |
| D100 | 0.02611(24) | 0.0145(4) | 0.1392(4) | 0.5 | 0.0201(7) |
| D101 | 0.46895(24) | 0.9925(6) | 0.0790(5) | 0.5 | 0.0201(7) |
| D200 | 0.34980(25) | 0.1021(6) | 0.0414(13) | 0.5 | 0.0201(7) |
| D201 | 0.3875(8) | 0.33831(21) | 0.2769(16) | 0.5 | 0.0201(7) |
| D202 | 0.1047(8) | 0.41917(31) | 0.4079(7) | 0.5 | 0.0201(7) |
| D203 | 0.42248(35) | 0.1225(7) | 0.0927(7) | 0.5 | 0.0201(7) |
| D300 | 0.22739(20) | 0.0839(6) | 0.3647(24) | 0.5 | 0.0182(6) |
| D301 | 0.1564(4) | 0.0758(6) | 0.2522(7) | 0.5 | 0.0182(6) |
| D302 | 0.1533(4) | 0.04809(25) | 0.3516(23) | 0.5 | 0.0182(6) |
| D303 | 0.1638(5) | 0.13063(25) | 0.4031(13) | 0.5 | 0.0182(6) |
| D310 | 0.0535(5) | 0.22378(23) | 0.1048(17) | 0.5 | 0.0182(6) |
| D311 | 0.0610(7) | 0.1698(7) | 0.2481(6) | 0.5 | 0.0182(6) |
| D312 | 0.03651(32) | 0.1546(4) | 0.0637(10) | 0.5 | 0.0182(6) |
| D313 | 0.11487(23) | 0.1820(8) | 0.0868(13) | 0.5 | 0.0182(6) |
| D400 | 0.30951(21) | 0.0695(7) | 0.0136(18) | 0.5 | 0.0182(6) |
| D401 | 0.2470(6) | 0.4300(5) | 0.3426(6) | 0.5 | 0.0182(6) |
| D402 | 0.2551(8) | 0.03485(24) | 0.0696(13) | 0.5 | 0.0182(6) |
| D403 | 0.2409(5) | 0.11614(29) | 0.0805(11) | 0.5 | 0.0182(6) |
| D410 | 0.0596(5) | 0.31259(23) | 0.4395(18) | 0.5 | 0.0182(6) |
| D411 | 0.4694(6) | 0.2439(7) | 0.0670(6) | 0.5 | 0.0182(6) |
| D412 | 0.00504(29) | 0.2630(8) | 0.3701(9) | 0.5 | 0.0182(6) |
| D413 | 0.08077(27) | 0.2409(4) | 0.4270(19) | 0.5 | 0.0182(6) |
| D500 | 0.07610(27) | 0.0189(5) | 0.1218(11) | 0.5 | 0.0201(7) |
| D501 | 0.1174(8) | 0.4398(4) | 0.2933(6) | 0.5 | 0.0201(7) |
| D502 | 0.14689(32) | 0.0544(5) | 0.1431(8) | 0.5 | 0.0201(7) |
| D503 | 0.0184(7) | 0.1359(8) | 0.9894(21) | 0.5 | 0.0201(7) |
| D510 | 0.0058(6) | 0.07260(35) | 0.3925(8) | 0.5 | 0.0201(7) |
| D511 | 0.44778(34) | 0.1095(8) | 0.1769(7) | 0.5 | 0.0201(7) |
| D512 | 0.0323(7) | 0.1253(4) | 0.3730(7) | 0.5 | 0.0201(7) |
| D513 | 0.3783(5) | 0.02203(24) | 0.2572(17) | 0.5 | 0.0201(7) |
| D600 | 0.1888(6) | 0.21556(27) | 0.1087(11) | 0.5 | 0.0390(15) |
| D601 | 0.2288(4) | 0.13554(26) | 0.0517(28) | 0.5 | 0.0390(15) |
| D602 | 0.15792(23) | 0.1661(7) | 0.0111(14) | 0.5 | 0.0390(15) |
| D603 | 0.2109(7) | 0.2958(4) | 0.3043(7) | 0.5 | 0.0390(15) |
| D610 | 0.3183(8) | 0.2625(4) | 0.0513(7) | 0.5 | 0.0390(15) |
| D611 | 0.1844(8) | 0.2352(4) | 0.3578(7) | 0.5 | 0.0390(15) |
| D612 | 0.36704(26) | 0.2682(4) | -0.0184(19) | 0.5 | 0.0390(15) |
| D613 | 0.3124(6) | 0.16184(22) | 0.3101(20) | 0.5 | 0.0390(15) |

Space group: $I\bar{4}2d$, Lattice parameters: $a = b = 19.74779(22)$ Å, $c = 7.73916(12)$ Å.

## Extended Data Table 2 | Fractional coordinates, site occupancies, and isotropic atomic displacement parameters ($U_{iso}$) for ice XXII at 1.48 GPa, 93 K.

| Site | $x$ | $y$ | $z$ | Occupancy | $U_{iso}$ | Site | $x$ | $y$ | $z$ | Occupancy | $U_{iso}$ |
|---|---|---|---|---|---|---|---|---|---|---|---|
| O1 | 0.17408(33) | 0.06836(34) | 0.2234(13) | 1 | 0.0237(9) | D90 | 0.1620(5) | 0.0092(6) | 0.48401(21) | 0.5 | 0.00557(35) |
| O2 | 0.33826(30) | 0.0903(4) | 0.18952(14) | 1 | 0.0237(9) | D91 | 0.5815(4) | 0.2260(24) | 0.25323(19) | 0.5 | 0.00557(35) |
| O3 | 0.4288(4) | 0.04078(34) | 0.07378(13) | 1 | 0.0237(9) | D92 | 0.6877(5) | 0.2077(10) | 0.21522(18) | 0.5 | 0.00557(35) |
| O4 | 0 | 0 | 0.27867(11) | 1 | 0.0237(9) | D93 | 0.3994(13) | 0.0321(16) | 0.02405(15) | 0.5 | 0.00557(35) |
| O5 | 0.11993(31) | 0.2280(4) | 0.15989(13) | 1 | 0.0237(9) | D100 | 0.1386(7) | 0.0075(9) | 0.17265(17) | 0.5 | 0.00557(35) |
| O6 | 0.0395(4) | 0.12207(27) | 0.07768(15) | 1 | 0.0237(9) | D101 | 0.0836(12) | 0.0329(4) | 0.12584(19) | 0.5 | 0.00557(35) |
| O7 | 0 | 0 | 0.00270(14) | 1 | 0.0237(9) | D102 | 0.1911(5) | 0.1960(5) | 0.40291(20) | 0.5 | 0.00557(35) |
| O8 | 0.51630(34) | 0.16854(26) | 0.13219(16) | 1 | 0.0237(9) | D103 | 0.1362(20) | 0.2012(6) | 0.36485(17) | 0.5 | 0.00557(35) |
| O9 | 0.63935(27) | 0.2250(5) | 0.23716(14) | 1 | 0.0237(9) | D110 | 0.2917(8) | 0.1392(9) | 0.12883(17) | 0.5 | 0.00557(35) |
| O10 | 0.1033(4) | -0.01438(28) | 0.14458(14) | 1 | 0.0237(9) | D111 | 0.2063(5) | 0.1876(8) | 0.11234(20) | 0.5 | 0.00557(35) |
| O11 | 0.25930(34) | 0.16151(32) | 0.09961(14) | 1 | 0.0237(9) | D112 | 0.2927(7) | 0.2051(5) | 0.08427(19) | 0.5 | 0.00557(35) |
| O12 | 0.11455(29) | 0.2677(4) | 0.05060(15) | 1 | 0.0237(9) | D113 | 0.2398(12) | 0.1278(6) | 0.07131(17) | 0.5 | 0.00557(35) |
| O13 | 0.13012(26) | 0.0426(4) | 0.34837(16) | 1 | 0.0237(9) | D120 | 0.0817(7) | 0.2239(4) | 0.06613(19) | 0.5 | 0.00557(35) |
| O14 | 0.2314(5) | 0.13524(26) | 0.45118(13) | 1 | 0.0237(9) | D121 | 0.1624(5) | 0.2877(10) | 0.07180(18) | 0.5 | 0.00557(35) |
| O15 | 0.2238(4) | 0.08209(28) | 0.00838(14) | 1 | 0.0237(9) | D122 | 0.0542(4) | 0.2841(10) | 0.04430(20) | 0.5 | 0.00557(35) |
| O16 | 0.0494(4) | 0.16880(33) | 0.51296(13) | 1 | 0.0237(9) | D123 | 0.1255(19) | 0.2928(8) | 0.01732(16) | 0.5 | 0.00557(35) |
| O17 | 0.08728(30) | 0.19186(31) | 0.29805(15) | 1 | 0.0237(9) | D130 | 0.0741(4) | 0.0288(13) | 0.33272(19) | 0.5 | 0.00557(35) |
| O18 | 0.15849(31) | 0.08182(34) | 0.57844(14) | 1 | 0.0237(9) | D131 | 0.1721(6) | 0.0722(7) | 0.32727(19) | 0.5 | 0.00557(35) |
| O19 | 0.08084(34) | 0.54932(29) | 0.19191(15) | 1 | 0.0237(9) | D132 | 0.5857(7) | 0.2550(5) | 0.11878(19) | 0.5 | 0.00557(35) |
| O20 | 0.0078(4) | 0.1201(4) | 0.41255(12) | 1 | 0.0237(9) | D133 | 0.0759(5) | 0.0643(9) | 0.36278(20) | 0.5 | 0.00557(35) |
| D10 | 0.2357(4) | 0.0692(29) | 0.21308(19) | 0.5 | 0.00557(35) | D140 | 0.2406(20) | 0.1693(6) | 0.48123(17) | 0.5 | 0.00557(35) |
| D11 | 0.1449(9) | 0.0445(8) | 0.19384(14) | 0.5 | 0.00557(35) | D141 | 0.0131(7) | 0.1561(5) | 0.18237(19) | 0.5 | 0.00557(35) |
| D12 | 0.1725(29) | 0.0234(5) | 0.24716(15) | 0.5 | 0.00557(35) | D142 | 0.1999(8) | 0.0862(4) | 0.45898(20) | 0.5 | 0.00557(35) |
| D13 | 0.1442(8) | 0.0984(7) | 0.25043(15) | 0.5 | 0.00557(35) | D143 | 0.2286(22) | 0.1669(6) | 0.41974(16) | 0.5 | 0.00557(35) |
| D20 | 0.2838(4) | 0.0749(12) | 0.20662(19) | 0.5 | 0.00557(35) | D150 | 0.0589(7) | 0.3638(6) | 0.23337(18) | 0.5 | 0.00557(35) |
| D21 | 0.3143(10) | 0.1130(9) | 0.15780(17) | 0.5 | 0.00557(35) | D151 | 0.2748(5) | 0.0479(6) | 0.01347(20) | 0.5 | 0.00557(35) |
| D22 | 0.3864(5) | 0.1071(11) | 0.21210(18) | 0.5 | 0.00557(35) | D152 | 0.2356(17) | 0.1149(6) | 0.03855(17) | 0.5 | 0.00557(35) |
| D23 | 0.3704(8) | 0.0417(4) | 0.18062(20) | 0.5 | 0.00557(35) | D153 | 0.2025(10) | 0.0372(4) | 0.02888(19) | 0.5 | 0.00557(35) |
| D30 | 0.4838(4) | 0.0145(7) | 0.06493(20) | 0.5 | 0.00557(35) | D160 | 0.0716(11) | 0.2198(4) | 0.50059(20) | 0.5 | 0.00557(35) |
| D31 | 0.0606(17) | 0.4165(5) | 0.09844(18) | 0.5 | 0.00557(35) | D161 | 0.0055(4) | 0.3143(11) | 0.02888(19) | 0.5 | 0.00557(35) |
| D32 | 0.4088(10) | 0.0407(19) | 0.03811(15) | 0.5 | 0.00557(35) | D162 | 0.0746(9) | 0.1437(8) | 0.54342(17) | 0.5 | 0.00557(35) |
| D33 | 0.1009(8) | 0.4933(6) | 0.09860(18) | 0.5 | 0.00557(35) | D163 | 0.0356(16) | 0.1474(9) | 0.47894(16) | 0.5 | 0.00557(35) |
| D40 | 0.0329(5) | 0.0079(16) | 0.31046(14) | 0.5 | 0.00557(35) | D170 | 0.1208(7) | 0.1547(5) | 0.27689(18) | 0.5 | 0.00557(35) |
| D41 | 0.2457(32) | 0.2297(7) | 0.49361(13) | 0.5 | 0.00557(35) | D171 | 0.1239(6) | 0.1920(29) | 0.32888(16) | 0.5 | 0.00557(35) |
| D52 | 0.1532(7) | 0.2350(21) | 0.19177(16) | 0.5 | 0.00557(35) | D172 | 0.0630(10) | 0.2449(4) | 0.29199(20) | 0.5 | 0.00557(35) |
| D53 | 0.0813(6) | 0.2602(6) | 0.13841(18) | 0.5 | 0.00557(35) | D173 | 0.0262(4) | 0.1778(13) | 0.30428(20) | 0.5 | 0.00557(35) |
| D50 | 0.1530(7) | 0.2177(16) | 0.12842(17) | 0.5 | 0.00557(35) | D180 | 0.1336(9) | 0.5298(4) | 0.08664(20) | 0.5 | 0.00557(35) |
| D51 | 0.0683(5) | 0.1948(6) | 0.16591(20) | 0.5 | 0.00557(35) | D181 | 0.1085(5) | 0.1064(8) | 0.56104(19) | 0.5 | 0.00557(35) |
| D62 | 0.0285(18) | 0.0940(7) | 0.04540(16) | 0.5 | 0.00557(35) | D182 | 0.2163(4) | 0.0727(18) | 0.56356(19) | 0.5 | 0.00557(35) |
| D63 | 0.0582(12) | 0.0783(5) | 0.09978(18) | 0.5 | 0.00557(35) | D183 | 0.1724(14) | 0.5302(9) | 0.11436(16) | 0.5 | 0.00557(35) |
| D60 | 0.0681(8) | 0.1731(4) | 0.07090(20) | 0.5 | 0.00557(35) | D190 | 0.3885(8) | -0.0045(4) | 0.17607(19) | 0.5 | 0.00557(35) |
| D61 | 0.2302(4) | 0.1072(9) | 0.32302(21) | 0.5 | 0.00557(35) | D191 | 0.5607(11) | 0.0977(4) | 0.17477(19) | 0.5 | 0.00557(35) |
| D72 | 0.20073(35) | 0.2425(16) | 0.22919(16) | 0.5 | 0.00557(35) | D192 | 0.1124(7) | 0.5500(28) | 0.22463(16) | 0.5 | 0.00557(35) |
| D73 | 0.0261(7) | 0.04090(35) | 0.02451(17) | 0.5 | 0.00557(35) | D193 | 0.0264(4) | 0.5226(7) | 0.20180(20) | 0.5 | 0.00557(35) |
| D80 | 0.0220(6) | 0.3692(5) | 0.11460(19) | 0.5 | 0.00557(35) | D200 | 0.2406(12) | 0.6622(6) | 0.19215(17) | 0.5 | 0.00557(35) |
| D81 | 0.0364(5) | 0.3020(7) | 0.14190(20) | 0.5 | 0.00557(35) | D201 | 0.0364(8) | 0.0930(7) | 0.38363(17) | 0.5 | 0.00557(35) |
| D82 | 0.5491(7) | 0.2138(4) | 0.11830(20) | 0.5 | 0.00557(35) | D202 | 0.0236(14) | 0.1494(7) | 0.44361(16) | 0.5 | 0.00557(35) |
| D83 | 0.5313(14) | 0.1251(5) | 0.15567(18) | 0.5 | 0.00557(35) | D203 | 0.2116(5) | 0.6071(9) | 0.14108(18) | 0.5 | 0.00557(35) |

Space group: $Fdd2$, Lattice parameters: $a$ = 14.9690(5) Å, $b$ = 16.4707(6) Å, $c$ = 25.4757(9) Å.

**Extended Data Table 3 | Fractional coordinates, site occupancies, and isotropic atomic displacement parameters ($U_{iso}$) for ice XXIII at 2.40 GPa, 93 K.**

| Site | x | y | z | Occupancy | $U_{iso}$ |
|---|---|---|---|---|---|
| O100 | 0.9965 | 0.7505 | 0.8615 | 1 | 0.0114 |
| O101 | 0.9983 | 0.7507 | 0.3874 | 1 | 0.0114 |
| O200 | 0.3935 | 0.8633 | 0.6262 | 1 | 0.0114 |
| O201 | 0.6092 | 0.6350 | 0.6200 | 1 | 0.0114 |
| O202 | 0.1142 | 0.3576 | 0.6297 | 1 | 0.0114 |
| O203 | 0.8861 | 0.1409 | 0.6246 | 1 | 0.0114 |
| O300 | 0.1792 | 0.8377 | 0.9796 | 1 | 0.0114 |
| O301 | 0.8194 | 0.6634 | 0.9787 | 1 | 0.0114 |
| O302 | 0.0879 | 0.5706 | 0.2705 | 1 | 0.0114 |
| O303 | 0.9115 | 0.9299 | 0.2717 | 1 | 0.0114 |
| O310 | 0.0699 | 0.9287 | 0.7506 | 1 | 0.0114 |
| O311 | 0.9312 | 0.5710 | 0.7530 | 1 | 0.0114 |
| O312 | 0.1779 | 0.6807 | 0.4961 | 1 | 0.0114 |
| O313 | 0.8185 | 0.8190 | 0.4964 | 1 | 0.0114 |
| O400 | 0.2614 | 0.8292 | 0.6462 | 1 | 0.0114 |
| O401 | 0.7385 | 0.6715 | 0.6458 | 1 | 0.0114 |
| O402 | 0.0787 | 0.4883 | 0.6057 | 1 | 0.0114 |
| O403 | 0.9214 | 0.0112 | 0.6035 | 1 | 0.0114 |
| O410 | 0.0438 | 0.0206 | 0.0744 | 1 | 0.0114 |
| O411 | 0.9587 | 0.4830 | 0.0751 | 1 | 0.0114 |
| O412 | 0.2677 | 0.7074 | 0.1753 | 1 | 0.0114 |
| O413 | 0.7308 | 0.7928 | 0.1768 | 1 | 0.0114 |
| O500 | 0.1164 | 0.7862 | 0.6868 | 1 | 0.0114 |
| O501 | 0.8828 | 0.7149 | 0.6857 | 1 | 0.0114 |
| O502 | 0.0355 | 0.6329 | 0.5645 | 1 | 0.0114 |
| O503 | 0.9643 | 0.8658 | 0.5631 | 1 | 0.0114 |
| O510 | 0.0224 | 0.8544 | 0.1048 | 1 | 0.0114 |
| O511 | 0.9778 | 0.6476 | 0.1017 | 1 | 0.0114 |
| O512 | 0.1037 | 0.7281 | 0.1473 | 1 | 0.0114 |
| O513 | 0.8982 | 0.7716 | 0.1430 | 1 | 0.0114 |
| O600 | 0.2034 | 0.9252 | 0.6674 | 1 | 0.0176 |
| O601 | 0.7976 | 0.5737 | 0.6690 | 1 | 0.0176 |
| O602 | 0.1759 | 0.5470 | 0.5776 | 1 | 0.0176 |
| O603 | 0.8246 | 0.9513 | 0.5815 | 1 | 0.0176 |
| O610 | 0.3251 | 0.9574 | 0.9179 | 1 | 0.0176 |
| O611 | 0.6748 | 0.5427 | 0.9192 | 1 | 0.0176 |
| O612 | 0.2078 | 0.4258 | 0.3309 | 1 | 0.0176 |
| O613 | 0.7922 | 0.0742 | 0.3299 | 1 | 0.0176 |
| D1001 | 0.9713 | 0.7305 | 0.7675 | 1 | 0.0071 |
| D1002 | 0.0033 | 0.7215 | 0.4869 | 0.511 | 0.0071 |
| D1003 | 0.9822 | 0.7770 | 0.4844 | 0.444 | 0.0071 |
| D1010 | 0.4690 | 0.7458 | 0.7119 | 0.818 | 0.0071 |
| D1011 | 0.5303 | 0.7399 | 0.7038 | 0.223 | 0.0071 |
| D1012 | 1.0154 | 0.2794 | 0.5423 | 1 | 0.0071 |
| D2000 | 0.3487 | 0.8741 | 0.6687 | 0.530 | 0.0071 |
| D2001 | 0.6498 | 0.6457 | 0.6845 | 0.547 | 0.0071 |
| D2002 | 0.0946 | 0.3991 | 0.5856 | 0.567 | 0.0071 |
| D2003 | 0.8744 | 0.0941 | 0.6011 | 0.441 | 0.0071 |
| D2010 | 0.3820 | 0.0879 | 0.9066 | 0.742 | 0.0071 |
| D2011 | 0.5967 | 0.4106 | 0.8962 | 0.092 | 0.0071 |
| D2012 | 0.3389 | 0.3486 | 0.3561 | 0.312 | 0.0071 |
| D2013 | 0.6615 | 0.1480 | 0.3411 | 0.635 | 0.0071 |
| D2020 | 0.1074 | 0.1688 | 0.0317 | 0.473 | 0.0071 |
| D2021 | 0.8941 | 0.3311 | 0.0301 | 0.550 | 0.0071 |
| D2022 | 0.4204 | 0.6302 | 0.2125 | 0.585 | 0.0071 |
| D2023 | 0.5827 | 0.8487 | 0.2170 | 0.478 | 0.0071 |
| D2030 | 0.4254 | 0.8735 | 0.7172 | 0.905 | 0.0071 |
| D2031 | 0.5785 | 0.6325 | 0.7168 | 0.161 | 0.0071 |
| D2032 | 0.1174 | 0.3267 | 0.5330 | 0.213 | 0.0071 |
| D2033 | 0.8803 | 0.1720 | 0.5291 | 0.769 | 0.0071 |
| D3000 | 0.2278 | 0.8318 | 0.9712 | 0.441 | 0.0071 |
| D3001 | 0.7731 | 0.6733 | 1.0100 | 0.559 | 0.0071 |
| D3002 | 0.0961 | 0.5224 | 0.2599 | 0.448 | 0.0071 |
| D3003 | 0.9138 | 0.9781 | 0.2490 | 0.552 | 0.0071 |
| D3010 | 0.1519 | 0.8366 | 0.8762 | 0.512 | 0.0071 |
| D3011 | 0.8482 | 0.6737 | 0.8813 | 0.481 | 0.0071 |
| D3012 | 0.0924 | 0.5998 | 0.3702 | 0.477 | 0.0071 |
| D3013 | 0.9162 | 0.9004 | 0.3707 | 0.558 | 0.0071 |
| D3020 | 0.1552 | 0.7969 | 1.0123 | 0.529 | 0.0071 |
| D3021 | 0.8453 | 0.7030 | 1.0119 | 0.438 | 0.0071 |
| D3022 | 0.0461 | 0.5942 | 0.2460 | 0.559 | 0.0071 |
| D3023 | 0.9500 | 0.9033 | 0.2344 | 0.364 | 0.0071 |
| D3030 | 0.1590 | 0.8790 | 0.0231 | 0.518 | 0.0071 |
| D3031 | 0.8341 | 0.6199 | 0.0228 | 0.522 | 0.0071 |
| D3032 | 0.1311 | 0.5864 | 0.2278 | 0.516 | 0.0071 |
| D3033 | 0.8716 | 0.9249 | 0.2005 | 0.527 | 0.0071 |
| D3100 | 0.0566 | 0.9739 | 0.7160 | 0.561 | 0.0071 |
| D3101 | 0.9470 | 0.5257 | 0.7274 | 0.439 | 0.0071 |
| D3102 | 0.2231 | 0.6973 | 0.5195 | 0.091 | 0.0071 |
| D3103 | 0.7747 | 0.8044 | 0.5378 | 0.909 | 0.0071 |
| D3110 | 0.0770 | 0.9168 | 0.8703 | 0.511 | 0.0071 |
| D3111 | 0.9455 | 0.5792 | 0.8703 | 0.522 | 0.0071 |
| D3112 | 0.1674 | 0.6963 | 0.3810 | 0.480 | 0.0071 |
| D3113 | 0.8271 | 0.8119 | 0.3743 | 0.531 | 0.0071 |
| D3120 | 0.0357 | 0.9036 | 0.6882 | 0.418 | 0.0071 |
| D3121 | 0.9644 | 0.6012 | 0.7029 | 0.534 | 0.0071 |
| D3122 | 0.1488 | 0.7164 | 0.5390 | 0.936 | 0.0071 |
| D3130 | 0.1126 | 0.9113 | 0.7082 | 0.510 | 0.0071 |
| D3131 | 0.8855 | 0.5881 | 0.7417 | 0.505 | 0.0071 |
| D3132 | 0.1615 | 0.6354 | 0.5206 | 0.494 | 0.0071 |
| D3133 | 0.8032 | 0.8642 | 0.5253 | 0.513 | 0.0071 |
| D4000 | 0.3081 | 0.8417 | 0.6657 | 0.470 | 0.0071 |
| D4001 | 0.6915 | 0.6841 | 0.6311 | 0.453 | 0.0071 |
| D4002 | 0.0617 | 0.4424 | 0.6139 | 0.433 | 0.0071 |
| D4003 | 0.9254 | 0.0579 | 0.5658 | 0.559 | 0.0071 |
| D4010 | 0.2455 | 0.1763 | 0.9714 | 0.569 | 0.0071 |
| D4011 | 0.7449 | 0.3160 | 0.9728 | 0.431 | 0.0071 |
| D4012 | 0.4329 | 0.5002 | 0.2788 | 0.566 | 0.0071 |
| D4013 | 0.5622 | 1.0036 | 0.2787 | 0.434 | 0.0071 |
| D4020 | 0.2413 | 0.7846 | 0.6578 | 0.614 | 0.0071 |
| D4021 | 0.7477 | 0.7140 | 0.7038 | 0.465 | 0.0071 |
| D4022 | 0.0347 | 0.4967 | 0.5554 | 0.752 | 0.0071 |
| D4023 | 0.9646 | -0.0038 | 0.5584 | 0.346 | 0.0071 |
| D4030 | 0.2413 | 0.8659 | 0.7114 | 0.485 | 0.0071 |
| D4031 | 0.7603 | 0.6328 | 0.6987 | 0.513 | 0.0071 |
| D4032 | 0.1164 | 0.5089 | 0.5456 | 0.381 | 0.0071 |
| D4033 | 0.8834 | 0.9947 | 0.5371 | 0.529 | 0.0071 |
| D4100 | 0.0656 | 0.0613 | 0.0327 | 0.908 | 0.0071 |
| D4101 | 0.9452 | 0.4369 | 0.1009 | 0.258 | 0.0071 |
| D4102 | 0.3124 | 0.6882 | 0.1605 | 0.365 | 0.0071 |
| D4103 | 0.6875 | 0.8126 | 0.2061 | 0.688 | 0.0071 |
| D4110 | 0.4661 | 0.9973 | 0.6878 | 0.437 | 0.0071 |
| D4111 | 0.5259 | 0.5046 | 0.6892 | 0.563 | 0.0071 |
| D4112 | 0.2495 | 0.2886 | 0.5589 | 0.571 | 0.0071 |
| D4113 | 0.7548 | 0.2157 | 0.5596 | 0.429 | 0.0071 |
| D4120 | 0.0063 | 0.0298 | 0.9974 | 0.249 | 0.0071 |
| D4121 | 0.9954 | 0.4882 | 0.9936 | 0.655 | 0.0071 |
| D4122 | 0.2536 | 0.7440 | 0.2501 | 0.535 | 0.0071 |
| D4123 | 0.7124 | 0.7575 | 0.2496 | 0.386 | 0.0071 |
| D4130 | 0.0812 | 0.9898 | 0.0939 | 0.436 | 0.0071 |
| D4131 | 0.9207 | 0.5115 | 0.0440 | 0.525 | 0.0071 |
| D4132 | 0.2378 | 0.6724 | 0.2189 | 0.529 | 0.0071 |
| D4133 | 0.7612 | 0.8305 | 0.1573 | 0.498 | 0.0071 |
| D5000 | 0.0752 | 0.7644 | 0.7253 | 1 | 0.0071 |
| D5002 | 0.0326 | 0.6756 | 0.5037 | 0.489 | 0.0071 |
| D5003 | 0.9829 | 0.8254 | 0.5106 | 0.543 | 0.0071 |
| D5010 | 0.1106 | 0.1867 | 0.9150 | 0.527 | 0.0071 |
| D5011 | 0.8702 | 0.3131 | 0.9119 | 0.450 | 0.0071 |
| D5012 | 0.4399 | 0.6179 | 0.3370 | 0.415 | 0.0071 |
| D5013 | 0.5618 | 0.8827 | 0.3379 | 0.522 | 0.0071 |
| D5020 | 0.1488 | 0.8040 | 0.7687 | 0.488 | 0.0071 |
| D5021 | 0.8538 | 0.6923 | 0.7683 | 0.519 | 0.0071 |
| D5022 | 0.0519 | 0.6017 | 0.4778 | 0.523 | 0.0071 |
| D5023 | 0.9384 | 0.8935 | 0.4841 | 0.442 | 0.0071 |
| D5030 | 0.0190 | 0.8853 | 0.6084 | 0.583 | 0.0071 |
| D5031 | 0.9810 | 0.6147 | 0.6084 | 0.466 | 0.0071 |
| D5032 | 0.8683 | 0.2417 | 0.8464 | 0.065 | 0.0071 |
| D5033 | 0.8707 | 0.7679 | 0.6229 | 1 | 0.0071 |
| D5101 | 0.9992 | 0.6789 | 0.0223 | 1 | 0.0071 |
| D5102 | 0.0693 | 0.7194 | 0.2331 | 0.778 | 0.0071 |
| D5103 | 0.9280 | 0.7537 | 0.2307 | 0.183 | 0.0071 |
| D5110 | 0.4476 | 0.0812 | 0.8029 | 0.094 | 0.0071 |
| D5111 | 0.5539 | 0.6418 | 0.8003 | 0.839 | 0.0071 |
| D5112 | 0.1122 | 0.3025 | 0.4503 | 0.786 | 0.0071 |
| D5113 | 0.8873 | 0.1972 | 0.4468 | 0.231 | 0.0071 |
| D5120 | 0.0339 | 0.8783 | 1.0000 | 0.489 | 0.0071 |
| D5121 | 0.9569 | 0.6283 | 1.0002 | 0.478 | 0.0071 |
| D5122 | 0.1275 | 0.7124 | 0.2486 | 0.520 | 0.0071 |
| D5123 | 0.8793 | 0.7838 | 0.2538 | 0.469 | 0.0071 |
| D5130 | 0.3746 | 0.7714 | 0.8496 | 0.562 | 0.0071 |
| D5131 | 0.6172 | 0.7293 | 0.9046 | 0.472 | 0.0071 |
| D5132 | 0.0210 | 0.3694 | 0.5816 | 0.636 | 0.0071 |
| D5133 | 0.9776 | 0.1278 | 0.4088 | 0.442 | 0.0071 |
| D6000 | 0.2087 | 0.9666 | 0.7329 | 0.435 | 0.0088 |
| D6001 | 0.8178 | 0.5331 | 0.7167 | 0.565 | 0.0088 |
| D6002 | 0.2162 | 0.5634 | 0.5203 | 0.444 | 0.0088 |
| D6003 | 0.7844 | 0.9364 | 0.5243 | 0.563 | 0.0088 |
| D6010 | 0.2264 | 0.8866 | 0.7175 | 0.515 | 0.0088 |
| D6011 | 0.7698 | 0.6136 | 0.6526 | 0.487 | 0.0088 |
| D6012 | 0.1375 | 0.5188 | 0.5486 | 0.619 | 0.0088 |
| D6013 | 0.8652 | 0.9784 | 0.5938 | 0.471 | 0.0088 |
| D6020 | 0.1591 | 0.9057 | 0.6869 | 0.490 | 0.0088 |
| D6021 | 0.8421 | 0.5924 | 0.6477 | 0.495 | 0.0088 |
| D6022 | 0.1583 | 0.5913 | 0.5477 | 0.506 | 0.0088 |
| D6023 | 0.8393 | 0.9090 | 0.6338 | 0.487 | 0.0088 |
| D6030 | 0.1856 | 0.0483 | 0.9288 | 0.453 | 0.0088 |
| D6031 | 0.7812 | 0.4495 | 0.9336 | 0.560 | 0.0088 |
| D6032 | 0.3011 | 0.5634 | 0.3157 | 0.486 | 0.0088 |
| D6033 | 0.7053 | 0.9574 | 0.3231 | 0.431 | 0.0088 |
| D6100 | 0.3193 | 0.0099 | 0.6725 | 0.377 | 0.0088 |
| D6101 | 0.6722 | 0.4896 | 0.6761 | 0.607 | 0.0088 |
| D6102 | 0.2609 | 0.4239 | 0.5745 | 0.532 | 0.0088 |
| D6103 | 0.7396 | 0.0600 | 0.5801 | 0.455 | 0.0088 |
| D6110 | 0.1830 | 0.9840 | 0.9781 | 0.547 | 0.0088 |
| D6111 | 0.8083 | 0.5181 | 0.9830 | 0.440 | 0.0088 |
| D6112 | 0.2305 | 0.5579 | 0.2725 | 0.514 | 0.0088 |
| D6113 | 0.7647 | 0.9354 | 0.2693 | 0.569 | 0.0088 |
| D6120 | 0.3676 | 0.0245 | 0.6228 | 0.594 | 0.0088 |
| D6121 | 0.6321 | 0.4774 | 0.6165 | 0.475 | 0.0088 |
| D6122 | 0.2743 | 0.3842 | 0.6243 | 0.471 | 0.0088 |
| D6123 | 0.7239 | 0.1190 | 0.6390 | 0.502 | 0.0088 |
| D6130 | 0.3124 | 0.9122 | 0.8813 | 0.478 | 0.0088 |
| D6131 | 0.6946 | 0.5870 | 0.9366 | 0.482 | 0.0088 |
| D6132 | 0.1611 | 0.4380 | 0.3531 | 0.473 | 0.0088 |
| D6133 | 0.8367 | 0.0591 | 0.2947 | 0.484 | 0.0088 |

Space group: $P2_12_12_1$, Lattice parameters: $a = 19.6973(23)$ Å, $b = 19.6482(20)$ Å, $c = 7.7361(7)$ Å.




\setcounter{page}{1}

  \maketitle

  {
    \footnotesize
    \noindent
    ${}^\textrm{*}$Corresponding author. E-mail: \href{mailto:hiroki@eqchem.s.u-tokyo.ac.jp}{hiroki@eqchem.s.u-tokyo.ac.jp}\; ORCiD: \href{https://orcid.org/0000-0002-3682-7558}{0000-0002-3682-7558}

    \noindent
    ${}^\textrm{\dag}$\textit{Now at} Institut für Anorganische Chemie, Fakultät für Chemie, Universität Wien, Währinger Straße 42, A-1090 Wien, Österreich
  }

\vspace{5mm}

\newpage
\tableofcontents
\newpage

\section{Crystal structure analysis}
\subsection{Overview of the structure analysis procedures for ices XXI and XXII}
\begin{enumerate}
  \item \textit{\textbf{Indexing using x-ray diffraction data.}} Peak search and indexing were performed using an autoindexing program \textit{CONOGRAPH} \cite{Oishi-Tomiyasu2014,Esmaeili2017}.
        This was successful for synchrotron x-ray diffraction patterns collected using imaging plate detectors (for diffraction measurements, we also used a flat-panel detector, but it yields worse signal-to-noise ratio and therefore patterns were not useful for indexing.
        For the detector choise, please refer to \reftab{tab:datasets}).
  \item  \textit{\textbf{Initial peak integration for x-ray diffraction data by the Le Bail or Pawley method.}}  Bragg-reflection intensities were extracted through whole-powder-pattern-fitting (WPPF) by the Le Bail method using the \textit{GSAS/EXPGUI} suite \cite{Larson2000,Toby2001}, or by the Pawley method using the \textit{Z-Rietveld} program \cite{Oishi2009,Oishi-Tomiyasu2012}.
  \item	\textit{\textbf{Improvement of the peak table employing the maximum entropy method.}} We improved the obtained reflection tables following a successful method proposed by Palatinus \textit{et al.}\ (2017) \cite{Palatinus2007a}. The maximum-entropy Patterson (MEP) method was adopted using the \textit{ERIS} program (available at \href{https://jp-minerals.org/eris/en/}{https://jp-minerals.org/eris/en/}, this is the successor program of \textit{Dysnomia} \cite{Momma2013}) for the reflection table from the peak extraction, \textit{i.e.}\ Patterson-function densities \[ \mathcal{P} (u,v,w) = \sum_{h,k,l} |F_{hkl}|^2 \exp \left( -2\pi i \left(hu+kv+lw \right) \right), \] where $u,v,w$ are real-space coordinates in the unit cell, were optimised according to the maximum entropy method (MEM) through numerical calculations of each $|F_{hkl}|^2$ value. The aim of MEP calculation is to (i) computationally estimate the best peak separation for the heavily overlapped peaks and (ii) estimate unobserved peak intensities at high-$Q$ regions, with a high physical likelihood. This does not require a phase recovery for structure factors (\textit{i.e.}\ MEP uses $|F_{hkl}|^2$, not $F_{hkl}$), making it executable before phase recovery attempts. Please refer to the manual of the \textit{ERIS} software for detailed algorithms. \\ 
  The resolution of grids in the unit cell was 0.2 \AA\  (\texttt{resolution 0.2}). Any reflections with $|I_\textrm{obs}| < 3\sigma(I)$ were treated as unobserved (\texttt{sigma\_cutoff 3}). Reflections whose $d$-value difference is smaller than 0.005 \AA\ were treated as overlapped, and therefore were grouped in the calculations (\texttt{delta\_d 0.005}). Calculations were performed at least until $R(I)<0.01$ and $R_\textrm{w} (I) < 0.01$ were achieved. All the other parameters were configured to be the default value. 
  \item \textit{\textbf{Elucidation of the oxygen-sublattice structures by the charge-flipping method.}} The improved reflection table was served for phase recovery by the charge flipping method using \textit{Superflip} software \cite{Palatinus2007}. The input files were generated automatically by the \textit{ERIS} suite, with some modifications by us for the charge flipping cycles. Reflections with a cumulative coverage of 90\% were used (using \texttt{reslimit}). \texttt{maxcycles}, \texttt{repeatmode}, and \texttt{bestdensities} options were adjusted to obtain reliable outputs; typical values are \texttt{maxcycles 50000–100000}, \texttt{repeatmode 100}, and \texttt{bestdensities 10} . For each cycle, we checked at least the top 10 candidates. We assigned oxygen atoms to each peak in the obtained electron densities and computed their powder x-ray diffraction patterns. Any candidates whose diffractogram completely differed from the experimental ones were rejected (we note this kind of case is very rare). This protocol resulted in the final structural candidates. We performed MEP and charge flipping calculations for several possible $Z$ numbers (number of molecules in the unit cell), and we evaluated each output by simulating the powder x-ray diffraction pattern and comparing it with experimental data. In both ices XXI and XXII, only one $Z$ number resulted in a good pattern in accord with the experiments.
  \item \textit{\textbf{Adding hydrogen atoms by molecular dynamics simulations.}} We used molecular dynamics (MD) simulations to determine the hydrogen-bonded network structures (\textit{i.e.}\ positions of the hydrogen atoms). MD simulations were performed until the potential energy was stabilised at temperatures similar to where neutron diffraction experiments were carried out. 
  \item \textit{\textbf{Constructing structural models for Rietveld fitting to neutron diffraction data.}} A crystallographic model (\textit{i.e.}\ structural model with a space group symmetry and the minimal number of atoms) was constructed using the time-averaged atomic distribution obtained from trajectories of MD simulations.
  \item  \textit{\textbf{Full structure refinement.}} Rietveld analysis was performed for experimental neutron diffraction profiles, starting from the structural model constructed in the last step. The \textit{GSAS/EXPGUI} suite was used. The deuterium site occupancies were constrained to follow the ice rules using the `constraints' and `restraints' functions in \textit{GSAS}.

\end{enumerate}

\noindent
\textit{Note.} 
The structure of ice XXIII, the low-temperature phase related to ice XXI, was analysed based on the structural model of ice XXI by considering group--subgroup relations. See Section \ref{subsec:ice_alpha_ordering}.

\subsection{Powder diffraction datasets}
Please refer to \reftab{tab:datasets} for the list of powder diffraction data we used for structure analyses and brief information about the sample composition, sample environment, data collection, pressure-temperature conditions, and crystallisation pathway.

{
  \footnotesize
  \fontencoding{T1}\fontfamily{phv}\selectfont 

  \begin{landscape}
    \begin{longtable}{lllllllllll}
\caption{List of powder diffraction data used for structure analysis}
\label{tab:datasets}\\
\multicolumn{11}{l}{\textbf{Ice XXI}} \\ \hline
\endfirsthead
\multicolumn{11}{c}%
{{\bfseries Table \thetable\ continued from previous page}} \\
\endhead
\multicolumn{1}{l}{\textbf{Data ID}} &
  \multicolumn{1}{l}{\textbf{\textit{T}   (K)}} &
  \multicolumn{1}{l}{\textbf{\textit{p}   (GPa)}} &
  \multicolumn{1}{l}{\textbf{\begin{tabular}[c]{@{}l@{}}Radiation\\ Source\end{tabular}}} &
  \multicolumn{1}{l}{\textbf{Instrumentation}} &
  \multicolumn{1}{l}{\textbf{Date}} &
  \multicolumn{1}{l}{\textbf{\begin{tabular}[c]{@{}l@{}}Water:matrix\\ volume ratio\end{tabular}}} &
  \multicolumn{1}{l}{\textbf{From}} &
  \multicolumn{1}{l}{\textbf{Other phase(s)}} &
  \multicolumn{1}{l}{\textbf{Used for}} &
  \multicolumn{1}{l}{\textbf{Main Text Figure}} \\ \hline
\multicolumn{1}{l}{Exp135-003} &
  \multicolumn{1}{l}{295} &
  \multicolumn{1}{l}{3.43} &
  \multicolumn{1}{l}{SXR} &
  \multicolumn{1}{l}{DAC, BL-18C, FPD} &
  \multicolumn{1}{l}{12 Nov. 2022} &
  \multicolumn{1}{l}{7:10} &
  \multicolumn{1}{l}{Liq.} &
  \multicolumn{1}{l}{Ice VII} &
  \multicolumn{1}{l}{Indexing, Rietveld (O)} &
  \multicolumn{1}{l}{} \\ 
\rowcolor[HTML]{F2F2F2} 
\multicolumn{1}{l}{\cellcolor[HTML]{F2F2F2}Exp137-013} &
  \multicolumn{1}{l}{\cellcolor[HTML]{F2F2F2}215} &
  \multicolumn{1}{l}{\cellcolor[HTML]{F2F2F2}2.41} &
  \multicolumn{1}{l}{\cellcolor[HTML]{F2F2F2}SXR} &
  \multicolumn{1}{l}{\cellcolor[HTML]{F2F2F2}DAC, BL-18C, FPD} &
  \multicolumn{1}{l}{\cellcolor[HTML]{F2F2F2}17 Nov. 2022} &
  \multicolumn{1}{l}{\cellcolor[HTML]{F2F2F2}7:10} &
  \multicolumn{1}{l}{\cellcolor[HTML]{F2F2F2}HDA} &
  \multicolumn{1}{l}{\cellcolor[HTML]{F2F2F2}none} &
  \multicolumn{1}{l}{\cellcolor[HTML]{F2F2F2}Indexing, Rietveld (O)} &
  \multicolumn{1}{l}{\cellcolor[HTML]{F2F2F2}} \\ 
\multicolumn{1}{l}{Exp177-004} &
  \multicolumn{1}{l}{295} &
  \multicolumn{1}{l}{$\sim$2.5} &
  \multicolumn{1}{l}{SXR} &
  \multicolumn{1}{l}{DAC, BL-18C, IP} &
  \multicolumn{1}{l}{8–9 Mar. 2023} &
  \multicolumn{1}{l}{2:10} &
  \multicolumn{1}{l}{Liq.} &
  \multicolumn{1}{l}{none} &
  \multicolumn{1}{l}{Indexing, Rietveld (O)} &
  \multicolumn{1}{l}{Fig 1d} \\ 
\rowcolor[HTML]{F2F2F2} 
\multicolumn{1}{l}{\cellcolor[HTML]{F2F2F2}Exp305-017} &
  \multicolumn{1}{l}{\cellcolor[HTML]{F2F2F2}215} &
  \multicolumn{1}{l}{\cellcolor[HTML]{F2F2F2}3.2} &
  \multicolumn{1}{l}{\cellcolor[HTML]{F2F2F2}SXR} &
  \multicolumn{1}{l}{\cellcolor[HTML]{F2F2F2}DAC, BL-18C, IP} &
  \multicolumn{1}{l}{\cellcolor[HTML]{F2F2F2}23 Mar. 2024} &
  \multicolumn{1}{l}{\cellcolor[HTML]{F2F2F2}7:10} &
  \multicolumn{1}{l}{\cellcolor[HTML]{F2F2F2}HDA} &
  \multicolumn{1}{l}{\cellcolor[HTML]{F2F2F2}Ice VII} &
  \multicolumn{1}{l}{\cellcolor[HTML]{F2F2F2}\begin{tabular}[c]{@{}l@{}}Indexing, LB-MEP-CF, \\ Rietveld (O)\end{tabular}} &
  \multicolumn{1}{l}{\cellcolor[HTML]{F2F2F2}} \\ 
\multicolumn{1}{l}{\begin{tabular}[c]{@{}l@{}}Exp150\_\\    HPN083259\end{tabular}} &
  \multicolumn{1}{l}{204} &
  \multicolumn{1}{l}{2.55} &
  \multicolumn{1}{l}{Neutron} &
  \multicolumn{1}{l}{Mito, PLANET} &
  \multicolumn{1}{l}{17 Dec. 2022} &
  \multicolumn{1}{l}{7:10} &
  \multicolumn{1}{l}{HDA} &
  \multicolumn{1}{l}{Pb} &
  \multicolumn{1}{l}{Rietveld (O+D)} &
  \multicolumn{1}{l}{Fig. 2e} \\ 

  \rowcolor[HTML]{F2F2F2}
  \multicolumn{1}{l}{\begin{tabular}[c]{@{}l@{}}Exp406:\\    Sample01-pos05-1\end{tabular}} &
  \multicolumn{1}{l}{90} &
  \multicolumn{1}{l}{0.0001} &
  \multicolumn{1}{l}{SXR} &
  \multicolumn{1}{l}{Rec, BL32XU, Eiger} &
  \multicolumn{1}{l}{29 Jan. 2025} &
  \multicolumn{1}{l}{5:10} &
  \multicolumn{1}{l}{HDA} &
  \multicolumn{1}{l}{none} &
  \multicolumn{1}{l}{Rietveld (O)} &
  \multicolumn{1}{l}{Extended Data Fig. 3a} \\ 
 &
   &
   &
   &
   &
   &
   &
   &
   &
   &
   \\ 
\multicolumn{11}{l}{\textbf{Ice XXIII: ordered   counterpart of ice XXI}}  \\ \hline
\multicolumn{1}{l}{\begin{tabular}[c]{@{}l@{}}Exp150\_\\    HPN083260\end{tabular}} &
  \multicolumn{1}{l}{93} &
  \multicolumn{1}{l}{2.40} &
  \multicolumn{1}{l}{Neutron} &
  \multicolumn{1}{l}{Mito,   PLANET} &
  \multicolumn{1}{l}{18   Dec. 2022} &
  \multicolumn{1}{l}{7:10} &
  \multicolumn{1}{l}{HDA} &
  \multicolumn{1}{l}{Pb} &
  \multicolumn{1}{l}{Rietveld (O+D)} &
  \multicolumn{1}{l}{Fig. 4e} \\ 
\multicolumn{11}{l}{} \\ 
\multicolumn{11}{l}{\textbf{Ice XXII}} \\ \hline
\multicolumn{1}{l}{Exp132-007} &
  \multicolumn{1}{l}{220} &
  \multicolumn{1}{l}{1.39} &
  \multicolumn{1}{l}{SXR} &
  \multicolumn{1}{l}{DAC, BL-18C, FPD} &
  \multicolumn{1}{l}{12 Oct. 2022} &
  \multicolumn{1}{l}{7:10} &
  \multicolumn{1}{l}{Liq.} &
  \multicolumn{1}{l}{Ice VI} &
  \multicolumn{1}{l}{Indexing} &
  \multicolumn{1}{l}{} \\ 
\rowcolor[HTML]{F2F2F2} 
\multicolumn{1}{l}{\cellcolor[HTML]{F2F2F2}Exp210-012} &
  \multicolumn{1}{l}{\cellcolor[HTML]{F2F2F2}250} &
  \multicolumn{1}{l}{\cellcolor[HTML]{F2F2F2}1.58} &
  \multicolumn{1}{l}{\cellcolor[HTML]{F2F2F2}SXR} &
  \multicolumn{1}{l}{\cellcolor[HTML]{F2F2F2}DAC, BL-18C, FPD} &
  \multicolumn{1}{l}{\cellcolor[HTML]{F2F2F2}9 May 2023} &
  \multicolumn{1}{l}{\cellcolor[HTML]{F2F2F2}2:10} &
  \multicolumn{1}{l}{\cellcolor[HTML]{F2F2F2}Liq.} &
  \multicolumn{1}{l}{\cellcolor[HTML]{F2F2F2}none} &
  \multicolumn{1}{l}{\cellcolor[HTML]{F2F2F2}Indexing} &
  \multicolumn{1}{l}{\cellcolor[HTML]{F2F2F2}} \\ 
\multicolumn{1}{l}{Exp304-013} &
  \multicolumn{1}{l}{250} &
  \multicolumn{1}{l}{1.65} &
  \multicolumn{1}{l}{SXR} &
  \multicolumn{1}{l}{DAC, BL-18C, FPD} &
  \multicolumn{1}{l}{20–21 Mar 2024} &
  \multicolumn{1}{l}{2:10} &
  \multicolumn{1}{l}{Liq.} &
  \multicolumn{1}{l}{Ice VI} &
  \multicolumn{1}{l}{Indexing} &
  \multicolumn{1}{l}{} \\ 
\rowcolor[HTML]{F2F2F2} 
\multicolumn{1}{l}{\cellcolor[HTML]{F2F2F2}Exp306-007} &
  \multicolumn{1}{l}{\cellcolor[HTML]{F2F2F2}230} &
  \multicolumn{1}{l}{\cellcolor[HTML]{F2F2F2}1.79} &
  \multicolumn{1}{l}{\cellcolor[HTML]{F2F2F2}SXR} &
  \multicolumn{1}{l}{\cellcolor[HTML]{F2F2F2}DAC, BL-18C, IP} &
  \multicolumn{1}{l}{\cellcolor[HTML]{F2F2F2}24 Mar 2024} &
  \multicolumn{1}{l}{\cellcolor[HTML]{F2F2F2}7:10} &
  \multicolumn{1}{l}{\cellcolor[HTML]{F2F2F2}Liq.} &
  \multicolumn{1}{l}{\cellcolor[HTML]{F2F2F2}Ice VI} &
  \multicolumn{1}{l}{\cellcolor[HTML]{F2F2F2}\begin{tabular}[c]{@{}l@{}}Indexing, P-MEP-CF, \\ Rietveld (O)\end{tabular}} &
  \multicolumn{1}{l}{\cellcolor[HTML]{F2F2F2}Fig. 1d} \\ 
\multicolumn{1}{l}{\begin{tabular}[c]{@{}l@{}}Exp220\_\\    HPN088031\end{tabular}} &
  \multicolumn{1}{l}{250} &
  \multicolumn{1}{l}{1.74} &
  \multicolumn{1}{l}{Neutron} &
  \multicolumn{1}{l}{Mito, PLANET} &
  \multicolumn{1}{l}{17 Apr. 2024} &
  \multicolumn{1}{l}{2:10} &
  \multicolumn{1}{l}{Liq.} &
  \multicolumn{1}{l}{Pb, ice VI} &
  \multicolumn{1}{l}{Rietveld   (O+D)} &
  \multicolumn{1}{l}{} \\ 
\rowcolor[HTML]{F2F2F2} 
\multicolumn{1}{l}{\cellcolor[HTML]{F2F2F2}\begin{tabular}[c]{@{}l@{}}Exp220\_\\    HPN088033\end{tabular}} &
  \multicolumn{1}{l}{\cellcolor[HTML]{F2F2F2}93} &
  \multicolumn{1}{l}{\cellcolor[HTML]{F2F2F2}1.48} &
  \multicolumn{1}{l}{\cellcolor[HTML]{F2F2F2}Neutron} &
  \multicolumn{1}{l}{\cellcolor[HTML]{F2F2F2}Mito, PLANET} &
  \multicolumn{1}{l}{\cellcolor[HTML]{F2F2F2}17 Apr. 2024} &
  \multicolumn{1}{l}{\cellcolor[HTML]{F2F2F2}2:10} &
  \multicolumn{1}{l}{\cellcolor[HTML]{F2F2F2}Liq.} &
  \multicolumn{1}{l}{\cellcolor[HTML]{F2F2F2}Pb, ice VI} &
  \multicolumn{1}{l}{\cellcolor[HTML]{F2F2F2}Rietveld (O+D)} &
  \multicolumn{1}{l}{\cellcolor[HTML]{F2F2F2}Fig. 3a} \\ 
\multicolumn{1}{l}{\begin{tabular}[c]{@{}l@{}}Exp408:\\    Sample02-pos03-1\end{tabular}} &
  \multicolumn{1}{l}{90} &
  \multicolumn{1}{l}{0.0001} &
  \multicolumn{1}{l}{SXR} &
  \multicolumn{1}{l}{Rec, BL32XU, Eiger} &
  \multicolumn{1}{l}{29 Jan. 2025} &
  \multicolumn{1}{l}{2:10} &
  \multicolumn{1}{l}{Liq.} &
  \multicolumn{1}{l}{Ice VI} &
  \multicolumn{1}{l}{\begin{tabular}[c]{@{}l@{}}P-MEP-CF, \\ Rietveld (O)\end{tabular}} &
  \multicolumn{1}{l}{Extended Data Fig. 3b} \\ \hline
\end{longtable}
\begin{tabular}{l}
  \textit{\textbf{Radiation source.}} SXR: synchrotron x-ray radiation monochromated at 20 keV; Neutron: spallation-source pulsed white neutrons\\
\textit{\textbf{High-pressure device.}} DAC: diamond anvil cell; Mito: Mito system (Komatsu \textit{et al.}, 2013 \cite{Komatsu2013}); Rec.: quench-recovered samples prepared using a piston cylinder apparatus \\
{\textit{\textbf{Detector (x-ray experiments).}}} FPD: flat-panel detector (CMOS); Eiger: DECTRIS Eiger X 9M, IP: imaging plate (Fujifilm, for medical imaging)\\
\textbf{\textit{Analysis.}} LB-MEP-CF: Le Bail fitting $\to$ MEP and charge flipping; P-MEP-CF: Pawley fitting $\to$ MEP and charge flipping; 
Rietveld (atom type): Rietveld analysis for specified element(s).

\end{tabular}
\end{landscape}
}

\clearpage
\subsection{Ice XXI}
\subsubsection{Unit cell}
A reliable indexing output was obtained from an in-situ x-ray diffraction pattern collected at 3.2 GPa, 210 K (\fbox{Exp305-017}).
It is noted that ice VII crystallises together with ice XXI from emulsified HDA when the pressure exceeds \textasciitilde3.0 GPa, and this is evident in another x-ray diffraction pattern of a sample crystallised from HDA at 2.5 GPa (\fbox{Exp135-003}) (for this experiment emulsion samples with a water:matrix ratio of 7:10 were used; reducing the amount of the aqueous phase may expand the pressure region in which single-phase ice XXI is obtained).
Because of the good data quality, especially in low-$Q$ regions, owing to a low-background imaging-plate detector, \fbox{Exp305-017} resulted in reliable indexing results even when another phase (ice VII) coexisted.
The unit cell of ice XXI was suggested as a top candidate when the obtained candidates were sorted using de Wolf's figure of merit with different $n$ values, $M_n$ ($n = 20, 30, 35, 40$) computed by the \textit{CONOGRAPH} suite.
We confirmed that other data, particularly high-quality synchrotron x-ray diffractograms \fbox{Exp137-013} (at 2.41 GPa, 215 K; no contaminant phases) and \fbox{Exp135-003} (at 2.5 GPa, 295 K; heavily contaminated with ice VII), can be indexed using this unit cell.

Based on the figure of merit, \textit{CONOGRAPH} suggested 17 leading candidates of space group listed in \reftab{tab:ice_alpha_SG_candidates}.
We were not able to further constrain the space group based on the reflection conditions (\textit{e.g.}\ judging the presence or absence of d-glide symmetry) due to very weak Bragg-reflection intensities at low-$Q$ regions. 
We discuss the space group based on structure refinement in the following steps. 

\begin{table*}[h]
  \caption{Space group candidates for ice XXI.}
  \label{tab:ice_alpha_SG_candidates}
  {
    \sffamily
    \begin{center}
      \begin{tabular}{|cccc|}
        \hline
        Space group            & \begin{tabular}{c}
                                   De Wolf's figure of merit, \\FoM ($M_{35}$)
                                 \end{tabular} & \begin{tabular}{c}
                                                   Can describe the \\O-sublattice   structure?
                                                 \end{tabular} & \begin{tabular}{c}
                                                                   Multiplicity of the \\general position
                                                                 \end{tabular}                           \\ \hline
        $ I4 $ (\#79)          & 8.617                                       & No                                           & 8  \\ 
        \rowcolor[HTML]{e6e6e6}
        { $ I\bar{4} $ (\#82) }   & 8.617                                       & Yes                                          & 8  \\ 
        $ I4/m $ (\#87)        & 8.617                                       & No                                           & 16 \\ 
        $ I422 $ (\#97)        & 8.617                                       & No                                           & 16 \\ 
        $ I4mm $ (\#107)       & 8.617                                       & No                                           & 16 \\ 
        $ I\bar{4}m2 $ (\#119) & 8.617                                       & No                                           & 16 \\ 
        $ I\bar{4}2m $ (\#121) & 8.617                                       & No                                           & 16 \\ 
        $ I4/mmm $ (\#139)     & 8.617                                       & No                                           & 16 \\ 
        $ I4_1md $ (\#109)     & 8.517                                       & No                                           & 16 \\ 
        \rowcolor[HTML]{f5ebeb}
        $ I\bar{4}2d $ (\#122) & 8.517                                       & Yes                                          & 16 \\ 
        \rowcolor[HTML]{e6e6e6}
        $ I4_1 $ (\#80)        & 8.147                                       & Yes                                          & 8  \\ 
        \rowcolor[HTML]{f5ebeb}
        $ I4_122 $ (\#98)      & 8.147                                       & Yes                                          & 16 \\ 
        \rowcolor[HTML]{f5ebeb}
        $ I4_1cd $ (\#110)     & 7.400                                       & Yes                                          & 16 \\ 
        $ I4cm $ (\#108)       & 7.302                                       & No                                           & 16 \\ 
        \rowcolor[HTML]{f5ebeb}
        $ I\bar{4}c2 $ (\#120) & 7.302                                       & Yes                                          & 16 \\ 
        $ I4/mcm $ (\#140)     & 7.302                                       & No                                           & 16 \\ \hline
      \end{tabular}%
    \end{center}
  }
\end{table*}

\subsubsection{Oxygen sublattice structure}

Next, each Bragg peak in x-ray diffraction patterns was integrated by the Le Bail method with the unit cell determined in the last step, by the \textit{GSAS} program on the \textit{EXPGUI} suite.
The primary x-ray diffraction data used for peak integration was \fbox{Exp137-013}.
We tested several phase recovery algorithms at this step (charge flipping, direct method, and real space method), but the set of extracted intensities at this step did not allow us to obtain a physically meaningful electron density. 

MEP calculations drastically improved the situations. Reflection intensities and estimated errors obtained by the Le Bail fitting were used as input data for MEP calculations. MEP cycles were run until $R$ values becomes below 1\%. 
The optimised values of $|F_{hkl}|^2$ were served for phase recovery by the charge flipping method using the \textit{Superflip} program.
Outputs from \textit{Superflip} were converted to a conventional crystallographic model by using the `Peak Search' function in \textit{VESTA}. We assigned oxygen atoms to electron clouds and simulated their powder x-ray diffraction pattern.
We performed the MEP$\to$charge-flipping calculations comprehensively for several possible $Z$ numbers that give a close density to ices VI and VII at the same pressure and temperature. 
We found the simulated x-ray diffraction pattern ($|F_{\textrm{calc}}|^2$) showed fairly good agreement with experimental data only when $Z = 152$. This yields a density (at 293 K, 2.4 GPa) between ices VI and VII.

We further refined the obtained oxygen-sublattice structure by using synchrotron powder x-ray diffraction profiles for the ``finalised'' version of oxygen-atom coordinates for the following steps. Additionally, this oxygen-sublattice structure was confirmed by Rietveld analyses performed for the \textit{ex-situ} data collected with a recovered sample (Extended Data Figure 3a).
Only small displacement of the oxygen atoms were observed, supporting tha validity of the output structure of the charge-flipping calculations.

It is noted that \textit{Superflip} includes the results of the space group search in the output files, and the most frequent space-group suggestion was $I\bar{4}c2$ [\#120] regardless of the input space group.
However, we noticed that other space groups can also describe the suggested oxygen-sublattice structures by allowing a small tolerance (less than 0.2 \AA) in oxygen-atom positions\footnote{
  We later found that these space groups were subgroups of $I4_1/acd$, the suggested space group for ice T2 by Yagasaki \textit{et al.} (2018) \cite{Yagasaki2018}.
}, as summarised in \reftab{tab:ice_alpha_SG_candidates}.
This reduced the number of space group candidates from 17 to 6.
We did not further constrain the space group here, because we ran symmetry-free MD simulations in the next step.
The input structure for MD simulations (in the $P1$ symmetry) was constructed from the $I\bar{4}c2$ model obtained by charge flipping followed by Rietveld analysis for x-ray data (\fbox{Exp137-013}). The choice of $I\bar{4}c2$ has little effects on the following discussions.
We finalise the space group symmetry later by the Rietveld analysis to neutron diffraction data.

\subsubsection{H-bonded network from MD simulations}

The preferential water orientations (\textit{i.e.}\ hydrogen-bonded network) in ice XXI were determined based on the oxygen positions obtained from x-ray diffraction data (\fbox{Exp137-013}).
The simulation box consisted of 1$\times$1$\times$3 of the unit cell, which contained 456 molecules.
First, flat-bottomed position restraints were imposed on the oxygen atoms, with a free radius of 0.05 nm and a harmonic force constant of 5000 kJ/mol/nm$^2$. 
Each water molecule was initially assigned a random orientation, and an NVT-MD simulation was performed at 200 K until the potential energy was stabilised. 
Subsequently, an NVT-MD simulation without the position restraints was conducted. 

To identify the most stable hydrogen-bond network, we iteratively performed NPT-MD simulations involving heating to 340~K and subsequent cooling back to 200~K, repeated twice at 2.5 GPa. 
The temperature was varied in 20 K increments, with each step simulated for 2 ns. 
The structure obtained at 200 K in the final cycle was thereafter energy-minimized using the steepest descent method.

To analyse MD simulation outputs in detail and to construct time-space averaged crystallographic structure models for Rietveld analysis in the next step, we used time-averaged atomic distribution --- instead of snapshots --- generated from the trajectory data of MD simulations.
The use of averaged atomic distributions minimise the effect of thermal motions in simulations, helping us to properly extract the averaged hydrogen-bonded network structures. 

In MD simulations of ice XXI, we found a potential-dependent structural feature around W6 molecules as mentioned in the main text (the labels for molecules from W1 to W6 are assigned to keep consistency with the model of ice T2 by Yagasaki \textit{et al.}) \cite{Yagasaki2018}.
The TIP4P/Ice potential suggested a structure with a normal tetrahedral network, in which there are four hydrogen sites around the O6 site (forming one W6–W3, one W6–W4, and two W6–W6 hydrogen bonds), whose occupancies are all 0.5. 
In contrast to the TIP5P suggestion described below, we call the structure of TIP4P/Ice `4-site model' hereafter. 

\begin{figure}
  \includegraphics{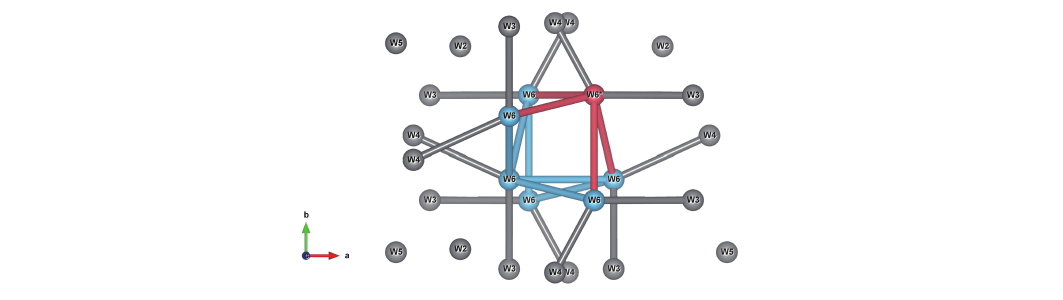}
  \caption{
    \textbf{6-site model constructed from time-averaged trajectory coordinates by TIP5P.} Balls and sticks represent oxygen atoms and hydrogen bonds, respectively. Bonds where W6 molecules are not involved are omitted.
  }
  \label{fig:alpha_6site_model}
\end{figure}

On the other hand, the averaged atomic distribution with the TIP5P potentials suggested that the W6–W6 network was more disordered.
Based on the time-averaged atomic distribution, we hereafter call this structure `6-site' structure.
The 6-site structure in the time-averaged picture is shown in \reffig{fig:alpha_6site_model}.
There are six hydrogen sites around the O6 site used for hydrogen bonds between other molecules.
If one focuses on a W6 molecule (the red ball labelled as `W6*' in \reffig{fig:alpha_6site_model}), there are one W3--W6 and one W4--W6 hydrogen bonds (these types of hydrogen bonds are found in every local structures around W6, gray sticks in \reffig{fig:alpha_6site_model}). Rest of the hydrogen bonds are formed with other W6 molecules (blue balls in \reffig{fig:alpha_6site_model}), with average probability of each W6--W6 hydrogen-bonds (red sticks in \reffig{fig:alpha_6site_model}) are 50\% in the time-averaged picture. Locally, no W6 molecules violate the ice rules. 

This is physically possible if the sum of the crystallographic occupancies of all the hydrogen sites around the O6 site is kept at 2: the hydrogen-disordered structure is described by the occupancies of hydrogen sites are 0.5 for those participating in W6–W3 and W6–W4 bonds and 0.25 for those consisting of the W6–W6 bonds.

It is noted that there were such ``additional'' W6--W6 bonds  even in the TIP4P/Ice calculations when we examined each snapshot in detail. However, such ``additional'' bonds are not a major structural feature in the TIP4P/Ice simulations, and therefore led to negligible atomic distributions there in time-averaged data. Trajectory analysis revealed such bonds are short-lived in TIP4P/Ice simulations, possessing a clear difference from TIP5P results. They are probably thermally induced since numbers of such bonds decreases with lowering temperature.

\subsubsection{Systematic Rietveld analysis for neutron diffraction data: constraining the network structure and space group}

In this step, we analysed neutron diffraction data to distinguish 4-site and 6-site structures suggested respectively by TIP4P/Ice and TIP5P, and to determine the space group symmetry of ice XXI.
We used \fbox{Exp150\_HPN083259} for systematic Rietveld analysis to determine the complete crystal structure including the hydrogen (deuterium) sublattice.
Unfortunately, Rietveld analysis readily resulted in divergence when the input space group was either $I\bar{4}$ or $I4_1$, \textit{i.e.}\ the space groups where the multiplicity of the general position is 8.
We need to use very strong dumping, constraints, and restraints for refinement with these two space groups, making unbiased comparisons with others impossible.
Thus, we only focus on the highest-symmetry space group candidates whose general position has a multiplicity of 16: $I\bar{4}2d$ [\#122], $I4_1 22$ [\#98], $I4_1 cd$ [\#110], and $I\bar{4}c2$ [\#120].
The choice of high-symmetry space groups is justified later because we succeeded in constructing a good high-symmetry model.

\begin{figure}
  \includegraphics{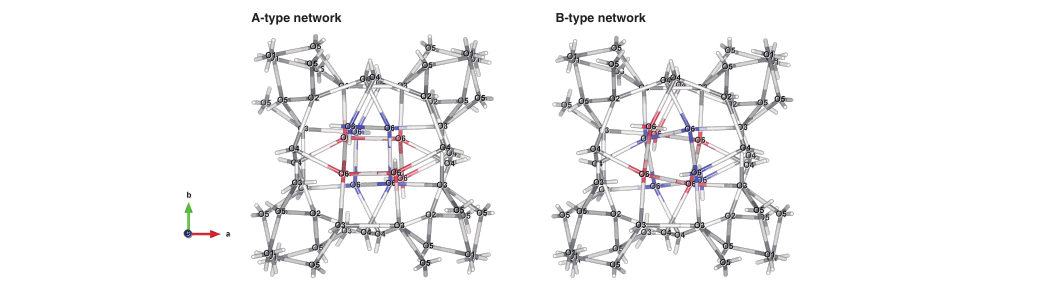}
  \caption{\textbf{Two types of possible W6–W6 network structures in the $I4_1 22$ 4-site models.}}
  \label{fig:ice_alpha_SG98_4site_models}
\end{figure}

Based on the symmetry operations, the candidate space groups are categorised into three groups:
\begin{enumerate}
  \item \underline{$I4_1 cd$ [\#110] and $I\bar{4}c2$ [\#120]} only accept 6-site models (suggested by TIP5P) due to symmetry operations.
  \item \underline{$I\bar{4}2d$ [\#122]} accepts one 4-site model (suggested by TIP4P/Ice) and one 6-site model (suggested by TIP5P).
  \item \underline{$I4_1 22$ [\#98]} accepts both 4-site and 6-site models:
  \begin{itemize}
    \item \textit{Any `4-site' structures that can be described with the $I4_1 22$ [\#98] space group symmetry were not observed in MD simulations}. Considering the symmetry operations of $I4_1 22$, there are two different types of W6–W6 network structures, namely network A and B shown in \reffig{fig:ice_alpha_SG98_4site_models}. In addition, there are two independent W6–W6 networks in the unit cell (called W60–W60 and W61–W61 networks, hereafter). Thus, there are three possible 4-site models in $I4_1 22$: the \textit{homo}-patterns where both W60–W60 and W61–W61 networks have a common network type (either A–A or B–B patterns), or the \textit{hetero}-pattern where W60–W60 and W61–W61 networks have different types of network structure (A–B). 
    \item Although none of A–A, A–B, and B–B 4-site models were observed in time-averaged atomic distribution generated from the trajectory data of MD simulations\footnote{
      These structure appeared only as local structures of the `6-site' structure.
    }, we tested these in Rietveld analysis just in case to validate theoretical results.
    \item The 6-site model is the 1:1 mixture of A–A and B–B networks. 
  \end{itemize}
\end{enumerate}

To compare all the possible structures, systematic Rietveld analyses were performed for $I4_1 cd$ (6-site model) and $I\bar{4}c2$ (6-site model), $I4_1 22$ (6-site and 4-site [A–A, A–B, and B–B patterns] models), and $I\bar{4}2d$ (6-site and 4-site models) using the \textit{GSAS} program for under the following conditions:
\begin{itemize}
  \item Lattice parameters were refined.
  \item Perfect hydrogen disorder was assumed. No deuterium-site occupancies were refined.
  \item 	`Distance Restraints' for the covalent-bonded O-D distances were configured to be 0.955 $\pm$ 0.03 \AA\ with a heavy `Restraint Weight' of 800. Looser constraints were also tested but refinement often resulted in unphysical molecular structures and/or divergence in some space groups\footnote{
    It is noted that calculations using $I\bar{4}2d$ were stable. This alone might imply the relevance of this space group.
  }.
  \item 	Isotropic atomic displacement parameters ($U_\textrm{iso}$) were used for all atoms. Using `Atom Constraints', two $U_\textrm{iso}$ values for deuterium atoms in W1–W5 molecules and W6 molecules were refined (\textit{i.e.}\ deuterium atoms in W1–W5 molecules have a common $U_\textrm{iso}$, and ones in W6 molecules have another common $U_\textrm{iso}$).
  \item 	The same background function was used for all models, as determined using the  `\textit{GSAS} background function 1' with the `Fit Background Graphically' function in the \textit{EXPGUI} software. Background parameters was not refined.
  \item 	To describe peak shapes,  `GSAS ToF Profile Function Type 3' was used. The `bet-0', `sig-1', and `gam-1' parameters were refined.
\end{itemize}

\begin{table}[]
  \caption{Summary of systematic Rietveld fitting for ice XXI against the neutron diffraction data (Exp150\_HPN083259)}
  \label{tab:ice_alpha_rietveld_sum}
  \begin{center}
    {
      \sffamily
      \begin{tabular}{|llllll|}
        \hline
        Space group                    & Model      & $\chi^2$    & $R_\textrm{wp}$ (\%) & $R_\textrm{p}$ (\%) & Notes                                                                                                                                                           \\ \hline
        \multirow{2}{*}{$I\bar{4}2d$ (\#122)} &
        6-site                         &
        3.316                          &
        1.68                          &
        1.55                           &
          Some D sites merged, yieliding a quasi-4-site model.\\
          & 4-site     & 2.670 & 1.54     & 1.48    &  \\ \hline
        \multirow{4}{*}{$I4_122$ (\#98)}  &
        6-site                         &
        (4.609)                        &
        (1.98)                         &
        (1.89)                         &

         Unphysical $U_\textrm{iso} (> 0.8)$ for some O atoms${}^\textrm{(*)}$.\\

                                       & 4-site [A--A] &  (4.753)  & (2.02) & (1.86) &  

         Unphysical $U_\textrm{iso} (> 0.8)$ for some O atoms${}^\textrm{(*)}$.\\
                                       & 4-site [A--B] &  (3.906)  &(1.82)&(1.67)&   Unphysical $U_\textrm{iso} (> 0.8)$ for some O atoms${}^\textrm{(*)}$.      \\ 
                                      
                                       & 4-site [B--B] &(4.491)&(1.95) &(1.79)  & Collapse of molecular structure of W6${}^\textrm{(**)}$.       \\ \hline
        $I4_1cd$ (\#110)                  & 6-site     & 9.857 & 2.88     & 2.47    &                                                                                                                                                                 \\ \hline
        $I\bar{4}c2$ (\#120)                  & 6-site     & 7.578 & 2.53     & 2.11    &                                                                                                                                                                 \\ \hline
      \end{tabular}
      
    }
  \end{center}
  {
    \sffamily \small
    \begin{tabular}{l}
          (*) Fixing the $U_\textrm{iso}$ values neither stabilised calculations nor improved fitting ($R_\textrm{wp} >$ \textasciitilde2.3\% and $\chi^2>$ \textasciitilde6 with a fixed $U_\textrm{iso}=$ 0.025).\\
          (**) Fixing/applying heavier dumping for deuterium positions of W6 induced the divergence of some $U_\textrm{iso}$ instead. 
      \end{tabular}
  }
  
\end{table}

\begin{figure*}
  \includegraphics{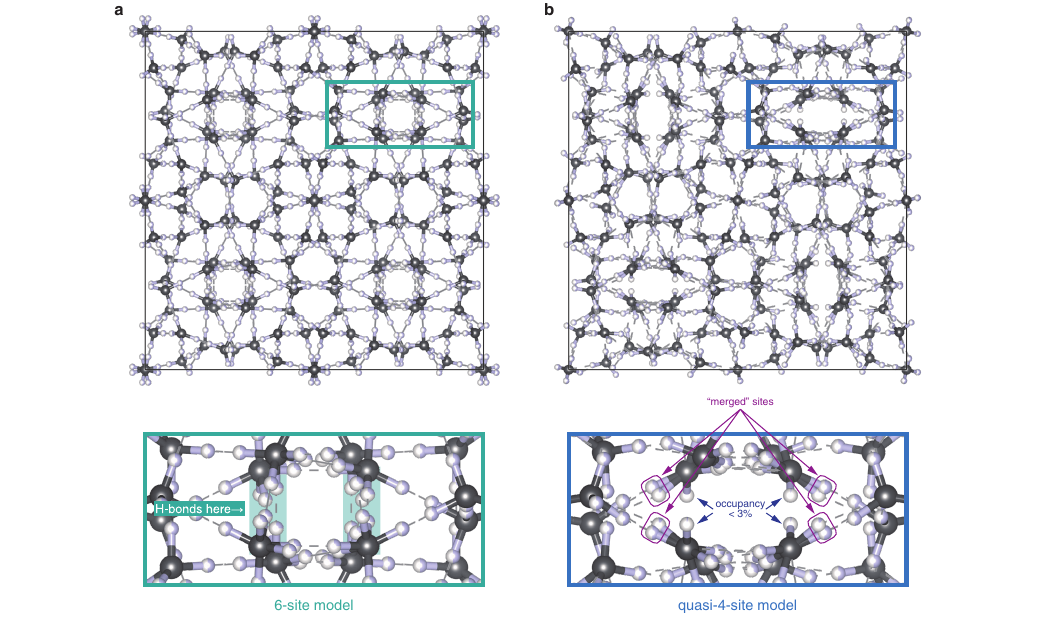}
  \caption{
    \textbf{Rietveld refinement using the 6-site model in the $I\bar{4}2d$ space group.}
    \textbf{a}, Input structure, constructed from time-averaged trajectory data of MD simulations using the TIP5P potential. 
    \textbf{b}, Refined structure (atomic coordinates, $U_\textrm{iso}$, and deuterium-site occupancies were refined). As shown in the enlarged figure, the refined structure no longer maintains the original ``6-site'' nature.
  }
  \label{fig:alpha_rietveld_sg122_6site}
\end{figure*}

Results of Rietveld analyses are summarised in \reftab{tab:ice_alpha_rietveld_sum}. 
The 4-site model structure in $I\bar{4}2d$ is the best model to fit the experimental neutron diffractograms. 
To double-check the 4-site model structure in $I\bar{4}2d$, we also performed Rietveld refinement for the 6-site model where the deuterium-site occupancies are refined under the constraints of the ice rule. 
The input and refined structures are summarised in \reffig{fig:alpha_rietveld_sg122_6site}.
Occupancies of some deuterium sites around the O6 site rapidly converged into values very close to 0 or slightly negative values, indicating that such hydrogen sites are ``unnecessary'' (\textit{cf}.\ In \reffig{fig:alpha_rietveld_sg122_6site}, there is an ``incorrect'' site with an occupancy of 3\%. This might be real due to thermally induced disorder, and might be the origin of the large $U_\textrm{iso}$ value for deuterium sites of W6 molecule in the finalised structure, but not significant enough to adopt the 6-site model against the 4-site model).
In addition, the similar trends, where ``wrong'' sites moves towards ``correct'' sites, were observed.
The resultant structure was a quasi-4-site model.

To conclude, the structure of ice XXI is well described by the 4-site model and the leading space group is $I\bar{4}2d$. 
Among the tested candidates, $I\bar{4}2d$ is the only space group that can describe the 4-site structure suggested by TIP4P/Ice force field in MD simulations. This suggests that TIP4P/Ice correctly reproduced the experimental structure of ice XXI whilst TIP5P failed.

\subsubsection{Finilising the $I\bar{4}2d$ model}

\subsubsection*{\underline{Atomic displacements in ice XXI: pronounced disorder around W6}}
Because refinement in $I\bar{4}2d$ was quite stable, we were able to loosen some constraints and restraints in the final cycles of refinement, yieliding the proposed model in the main text.
In particular, we loosened the constraints for $U_\textrm{iso}$ values referring to computational outputs as described below.
Initially, we introduced four $U_\textrm{iso}$ values for (i) O atoms of W1–W5, (ii) O atoms of W6, (iii) D atoms of W1–W5, and (iv) D atoms of D6. Nevertheless, we found an interesting results in our computational works. 
Computationally, displacement of each local molecule from the averaged position can be described by the root mean square displacement (RMSD) computed for each type of molecules W$_i$ using the coordinates of the centre of mass $\bm{r}_{\textrm{W}_i}$ as 
\[
  \rho^\textrm{RMSD} (\textrm{W}_k) = \sqrt{
    \frac{
      \sum_{i \in \textrm{W}_k} \Delta r^2_i
    }{
      N_{\textrm{W}_k}
    }
  },
\]
where
\[
  \Delta r^2 _i = \frac{
    \sum _t \left\{ \bm{r}_i (t) - \bar{\bm{r}}_i \right\}^2
  }{
    N_t
  }.
\]
The RMSD values for 10 types of molecules (W10, W20, W30, W31, W40, W41, W50, W51, W60, and W61) are summarised in \reftab{tab:ice_alpha_ADPs}.
These values are clearly categorised into three groups: small-displacement molecules (W10, W20, W50, W51), medium-displacement molecules (W30, W31, W40, W41), and large-displacement molecules (W60, W61). 
To check whether this is occurring in experiments, we refined six $U_\textrm{iso}$ (for O and D atoms for in the three types of molecules listed above) by Rietveld refinement for neutron diffraction data\footnote{
  We also tested refinement without any constraints about $U_\textrm{iso}$ but the results were unfortunate. The number of refined parameters was too large for Rietveld refinement and least-square calculation got unstable. 
}.
Refined $U_\textrm{iso}$ values are compared in \reftab{tab:ice_alpha_ADPs}. 
$U_\textrm{iso}$ for oxygen atoms exhibited the same trend as observed computationally, where O10, O20, O50, and O51 have the smallest, O30, O31, O40, and O41 have a medium, and O60 and O61 have the largest values of $U_\textrm{iso}$, respectively. $U_\textrm{iso}$ of D atoms of W60 and W61 molecules was significantly large, but differences were not significant between small-displacement molecules (W10, W20, W50, W51) and medium-displacement molecules (W30, W31, W40, W41).

\begin{figure}[H]
  \includegraphics{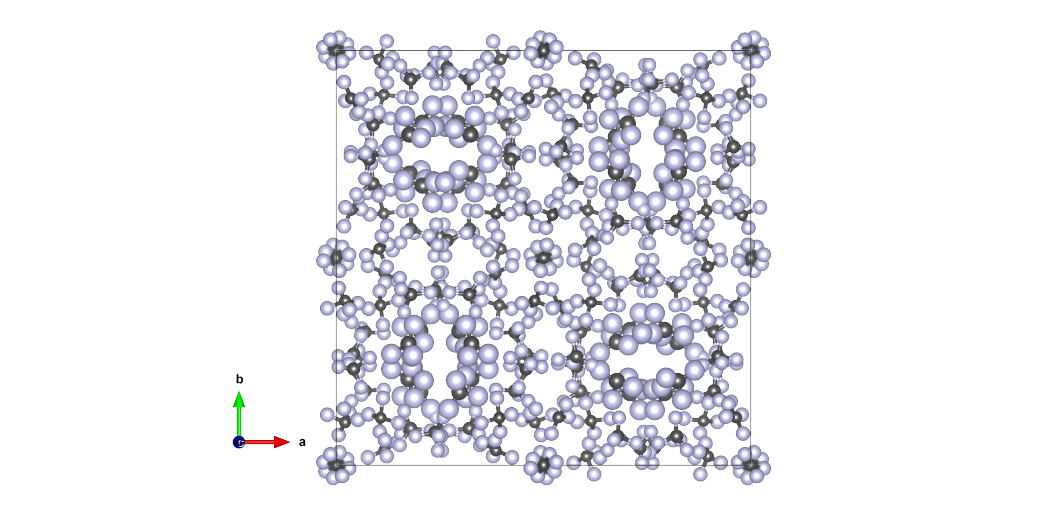}
  \caption{
    Refined structure of ice XXI. Balls represent the region where the probability of finding an atom is 90\% estimated base on the refined $U_\textrm{iso}$ values. 
  }
  \label{fig:ice_alpha_ADPs_illustration}
\end{figure}
\begin{table*}
  \caption{Atomic displacements in ice XXI by computational and experimental methods: RMSD (MD simulation) and $U_\textrm{iso}$ (Rietveld analysis, neutron diffraction) values}
  \label{tab:ice_alpha_ADPs}
  {
    \sffamily
    \begin{center}
      \begin{tabular}{|cccc|}
        \hline
      \multirow{2}{*}{Molecule} & Computational & \multicolumn{2}{c|}{Experimental}                             \\ 
      \cline{2-4} & $ \rho^\textrm{RMSD} $ (\AA)          & $U_\textrm{iso} (\textrm{O})$ & $U_\textrm{iso} (\textrm{D})$  \\ \hline
      W10& 0.201(4)&0.0131(6)&0.0201(7)\\
      W20& 0.201(4)&0.0131(6)&0.0201(7)            \\ 
      W30& 0.210(3)&0.0161(6)&0.0182(6)\\          
      W31&0.207(4)&0.0161(6)&0.0182(6)\\                  
      W40&0.207(4)&0.0161(6)&0.0182(6)\\          
      W41&0.209(4)&0.0161(6)&0.0182(6)\\            
      W50&0.202(4)&0.0131(6)&0.0201(7)\\          
      W51&0.202(3)&0.0131(6)&0.0201(7)\\            
      W60&0.222(3)&0.0241(14)&0.0390(15)\\          
      W61&0.229(3)&0.0241(14)&0.0390(15)\\    \hline        
      \end{tabular}
    \end{center}
  }
\end{table*}

These computational and experimental results revealed the pronounced positional disorder of W60 and W61 (W6 hereafter) molecules in ice XXI.
In the time-space averaged structure, refined using neutron diffraction data, the W6 molecules are heavily distorted with some $\angle$D-O-D $\approx$ 70\textdegree and the hydrogen-bonded network structure around W6 molecules displays bad tetrahedrality.
Thus the large $U_\textrm{iso}$ value in the averaged structure is likely to reflect the local positional displacement of the W6 molecules to improve the local tetrahedrality and to reduce local molecular distortions. 
Additinoally, we point out that some W6--W6 distances are longer than 3 \AA, which in general yields weak hydrogen bonds. Weak W6--W6 intereations also facilitates the large displacements of these molecules.
In \reffig{fig:ice_alpha_ADPs_illustration}, the refined time-space averaged structure of ice XXI is shown, with the sizes of balls (atoms) are adjusted to represent $xyz$ coordinates where the probability of finding atoms is 90\%.
Although atomic displacements of W6 molecules are quite large, we suggest that it is still feasible to maintain firm hydrogen-bond network structures around them. 
Related phenomena, such as rotational dynamics of the W6 molecules, would be nice subjects of the future studies.

\subsubsection*{\underline{\textit{Corroborating the orientationally disordered nature by refining the deuterium-site occupancies}}}
We checked whether our assumption of perfect hydrogen disorder is correct by refining deuterium-site occupancies. 
Deuterium site occupancies are not free variables but are constrained by Bernal–Fowler's ice rules [Sum of the deuterium-site occupancies around each oxygen site is 2, and sum of the deuterium-site occupancies on each hydrogen bond is 1]. 
In $I\bar{4}2d$, deuterium-site occupancies are described by 7 independent parameters, $o_i$ ($i \in [1,7]$) (called ordering parameters hereafter), whose initial value in the completely disordered structure is 0.5, as summarised in \reftab{tab:ice_alpha_ordering_SG122_expression}. 
These constraints on the site occupancies were implemented by using the `Constraints' function in \textit{GSAS/EXPGUI}. 
Refined ordering parameters and refinement statistics are summarised in \reftab{tab:ice_alpha_ordering_SG122_refinement}. 
The ordering parameters were somewhat altered from 0.5, and $R_\textrm{wp}$, $R_\textrm{p}$, and $\chi^2$ became slightly better. 
However, such a slight improvement is typical when the number of refined parameters is increased. No sites achieved >3\% ordering. 
This difference is insignificant, considering that many structural parameters are refined here. 
Consequently, we conclude that ice XXI is hydrogen disordered. 

\begin{table*}[]
  \caption{Expressing the deuterium-site occupancies in the $I\bar{4}2d$ model using seven ordering parameters.}
  \label{tab:ice_alpha_ordering_SG122_expression}

  {
    \sffamily
    \begin{center}
      \begin{tabular}{|cc|}
        \hline
        Site & Occupancy   representation \\ \hline
        D100 & $ o_1 $                    \\ 
        D101 & $ 1-o_1 $                  \\ 
        D200 & $ o_2 $                    \\ 
        D201 & $ o_3 $                    \\ 
        D202 & $ o_4 $                    \\ 
        D203 & $ 2-o_2-o_3-o_4 $          \\ 
        D300 & $ 0.5 $                    \\ 
        D301 & $ o_5 $                    \\ 
        D302 & $ o_6 $                    \\ 
        D303 & $ 1.5-o_5-o_6 $            \\ 
        D310 & $ 0.5 $                    \\ 
        D311 & $ o_1+o_2+o_3+o_4-o_6-1 $  \\ 
        D312 & $ 2-o_1-o_4-o_5 $          \\ 
        D313 & $ 0.5-o_2-o_3+o_5+o_6 $    \\ 
        D400 & $ 1-o_2 $                  \\ 
        D401 & $ 0.5 $                    \\ 
        D402 & $ o_7 $                    \\ 
        D403 & $ 0.5+o_2-o_7 $            \\ 
        D410 & $ 1-o_3 $                  \\ 
        D411 & $ 0.5 $                    \\ 
        D412 & $ 1-o_7 $                  \\ 
        D413 & $ o_3+o_7-0.5 $            \\ 
        D500 & $ 1-o_1 $                  \\ 
        D501 & $ 1-o_4 $                  \\ 
        D502 & $ 1-o_5 $                  \\ 
        D503 & $ o_1+o_4+o_5-1 $          \\ 
        D510 & $ o_1 $                    \\ 
        D511 & $ o_2+o_3+o_4-1 $          \\ 
        D512 & $ 2-o_1-o_2-o_3-o_4+o_6 $  \\ 
        D513 & $ 1-o_6 $                  \\ 
        D600 & $ 0.5 $                    \\ 
        D601 & $ 0.5-o_2+o_7 $            \\ 
        D602 & $ 0.5+o_2+o_3-o_5-o_6 $    \\ 
        D603 & $ 0.5-o_3+o_5+o_6-o_7 $    \\ 
        D610 & $ 0.5 $                    \\ 
        D611 & $ 0.5+o_3-o_5-o_6+o_7 $    \\ 
        D612 & $ 1.5-o_3-o_7 $            \\ 
        D613 & $ o_5+o_6-0.5 $            \\ \hline
      \end{tabular}
    \end{center}
  }
\end{table*}

\begin{table*}[]

  { \sffamily
    \begin{center}
      \caption{Refined ordering parameters, $R$ values and $\chi^2$ in fully disordered and partially ordered $I\bar{4}2d$ models for ice XXI.}
      \label{tab:ice_alpha_ordering_SG122_refinement}
      \begin{tabular}{|ccc|}
        \hline
                        & Fully disordered & Partially ordered \\ \hline
        $ o_1 $         & 0.5              & 0.490(6)         \\ 
        $ o_2 $         & 0.5              & 0.495(5)         \\ 
        $ o_3 $         & 0.5              & 0.499(5)         \\ 
        $ o_4 $         & 0.5              & 0.507(6)         \\ 
        $ o_5 $         & 0.5              & 0.507(5)         \\ 
        $ o_6 $         & 0.5              & 0.484(5)         \\ 
        $ o_7 $         & 0.5              & 0.521(6)         \\ 
        $R_\textrm{wp}$ & 1.15\%           & 1.13\%            \\ 
        $R_\textrm{p}$  & 1.02\%           & 1.01\%            \\ 
        $\chi^2$        & 1.922            & 1.798             \\ \hline
      \end{tabular}%

    \end{center}}
\end{table*}

\clearpage

\subsection{{Ice XXIII: low-temperature phase related to ice XXI}}
\label{subsec:ice_alpha_ordering}

We analysed the neutron diffraction profile of ice XXIII (\fbox{Exp150\_HPN083260}, cooled at \textasciitilde2.4 GPa at \textasciitilde0.2 K/min) based on the structure of ice XXI derived in the last section.
This low-temperature phase may be either a hydrogen-ordered phase or a disordered phase with some changes in the hydrogen-bond network structure as suggested for ice XIX \cite{Salzmann2021}. We took both possibilities into consideration and minimised our bias about the nature of ice XXIII. 

\begin{figure*}
  \begin{center}
    \includegraphics[width=1\linewidth]{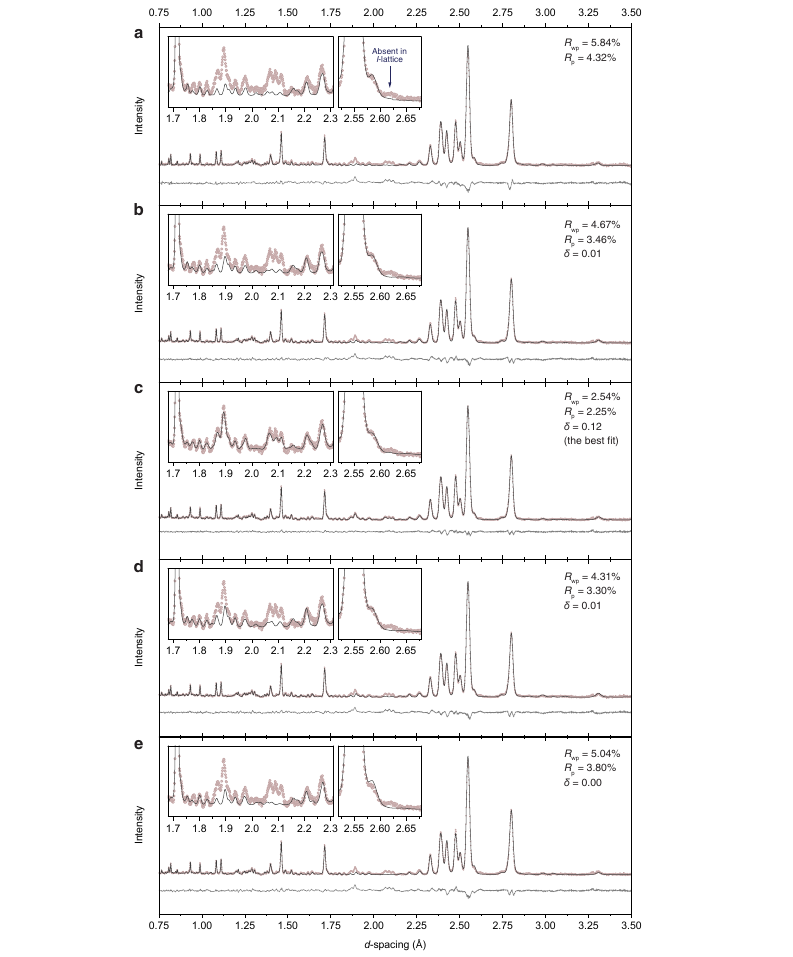}
  \end{center}
  \caption{
    \textbf{Observed, calculated, and difference neutron diffraction profiles for ice XXIII with several different space-group candidates.} 
    \textbf{a}, Completely disordered $I\bar{4}2d$ model as a reference (= ice XXI, deuterium-site occupancies were fixed at 0.5). 
    \textbf{b}, Partially ordered $P\bar{4}$ model. 
    \textbf{c}, Partially ordered $P2_12_12_1$ model (the best fit). 
      \textbf{d}, Partially ordered $P222_1$ [conjugacy class a] model.
      \textbf{e}, Partially ordered $P222_1$ [conjugacy class c] model.
    }
    \label{fig:ice_alpha_ordering_Rietveld}
\end{figure*}

We first focused on the positions of the new characteristic peaks of the low-temperature phase to constrain the space group.
Some new peaks are observed on cooling, and the positions of such new peaks cannot be explained by the body-centred tetragonal cell of ice XXI. 
In particular, we focused on a new peak at $d = 2.6177\; \textrm{\AA}$ (the peak indicated with an arrow in \reffig{fig:ice_alpha_ordering_Rietveld}a, inset).
This peak cannot be indexed by $I$- and $F$-lattices due to systematic extinction of Bragg reflections. 
This is a clear indication of symmetry reduction.

We tested the highest-symmetry tetragonal and orthorhombic subgroups that fulfil the reflection conditions for the new peak (\textit{i.e.}\ non-\textit{I} and non-\textit{F} lattices), as listed below. 
For $P222_1$, three models can be generated from $I\bar{4}2d$ structure, depending on the axis chosen to be the $2_1$ screw axis. They are defined as three `conjugacy classes' labelled as a, b, and c, following the practice in the Bilbao Crystallographic Server.

The space group candidates for ice XXIII are
  \begin{itemize}
  \item $P\bar{4}$: the unit cell is the same as $I\bar{4}2d$. This space group permits fully ordered structures.
  \item $P2_12_12_1$: the unit cell is transformed from $I\bar{4}2d$ by
  \[
    \textrm{rotation matrix:}
    \begin{pmatrix}
      1 & 0 & 0 \\
      0 & 1 & 0 \\
      0 & 0 & 1 \\
      \end{pmatrix},\;\;
      \textrm{origin shift:}
    \begin{pmatrix}
      0 \\
      1/4 \\
      3/8 \\
      \end{pmatrix} .
    \]
    This space group permits fully ordered structures.
  \item  $P222_1$ [conjugacy class a]:  the unit cell is transformed from $I\bar{4}2d$ by
  \[
    \textrm{rotation matrix:}
    \begin{pmatrix}
      0 & 0 & 1 \\
      1 & 0 & 0 \\
      0 & 1 & 0 \\
      \end{pmatrix},\;\;
      \textrm{origin shift:}
    \begin{pmatrix}
      1/4 \\
      1/2 \\
      3/8 \\
      \end{pmatrix} .
    \]
    This space group does not allow a fully hydrogen-ordered structure because the occupancies of D3000, D3001, D3102, D3103, D4010, D4011, D4112, D4113, D6002, D6003, D6102, and D6103 sites are fixed at 0.5 owing to the requirements from symmetry operations. 
    \item   $P222_1$ [conjugacy class b]: the unit cell is transformed from $I\bar{4}2d$ by
  \[
   \textrm{rotation matrix:}
    \begin{pmatrix}
      0 & 1 & 0 \\
      0 & 0 & 1 \\
      1 & 0 & 0 \\
      \end{pmatrix},\;\;
      \textrm{origin shift:}
    \begin{pmatrix}
      0 \\
      1/2 \\
      5/8 \\
      \end{pmatrix} .
    \]
      This space group is indistinguishable from $P222_1$ [conjugacy class a] on the basis of powder diffraction.
    \item $P222_1$ [conjugacy class c]: the unit cell is transformed from $I\bar{4}2d$ by
  \[
   \textrm{rotation matrix:}
    \begin{pmatrix}
      1 & 0 & 0 \\
      0 & 1 & 0 \\
      0 & 0 & 1 \\
      \end{pmatrix},\;\;
      \textrm{origin shift:}
    \begin{pmatrix}
      1/4 \\
      1/4 \\
      5/8 \\
      \end{pmatrix} .
    \]
    This space group does not allow a fully hydrogen-ordered structure because the occupancies of D3000, D3001, D3002, D3003, D3100, D3101, D3102, D3103, D4010, D4011, D4012, D4013, D4110, D4111, D4112, D4113, D6000, D6001, D6002, D6003, D6100, D6101, D6102, and D6013 sites are fixed at 0.5 owing to the requirements from symmetry operations. 
\end{itemize}

Each input structure for the ice XXIII was generated from the $I\bar{4}2d$ model of ice XXI by the \textit{VESTA} software \cite{Momma2011} through the following steps:
\begin{enumerate}
  \item  The symmetry-free $P1$ model was generated using the `Remove Symmetry' function in the `Unit Cell' tab of the `Edit Data' window. The `Update structure parameters to keep 3D geometry' option was used.
  \item If necessary, the lattice transformation matrix was specified using the `Unit Cell Transformation' subwindow in the `Unit Cell' tab of the `Edit Data' window\footnote{
    To examine the unit cell transformation, the `MAXSUB' function in the Bilbao Crystallographic Server \cite{Aroyo2006a,Aroyo2006} was used.
  }. The `Update structure parameters to keep 3D geometry' option was used.
  \item A target low-symmetry space group was chosen  in the `Unit Cell' tab of the `Edit Data' window. The `Keep structure parameters unchanged' option was used.
  \item `Initialize current matrix' was executed in the `Unit Cell Transformation' subwindow in the `Unit Cell' tab of the `Edit Data' window.
  \item `Remove duplicate atoms' was executed in the `Structure Parameters' tab of the `Edit Data' window.
\end{enumerate} 

For these highly complex structures, divergence of Rietveld analysis is unavoidable when atomic coordinates are refined simultaneously. 
Thus, we first refined the deuterium-site occupancies under the constraints of the ice rules (with fixed atomic coordinates) and thereafter the atomic coordinates with strong structural restraints about atomic distances (with fixed occupancies)\footnote{
    Due to this stepwise method, we used $R$ factors to evaluate Rietveld-fitting results, because $\chi^2$ values depend on the number of variable parameters. 

}. After that, the occupancies were refined again but only slight improvement was achieved.

We computed the $\delta$ index to evaluate the degree of hydrogen ordering as 
\[
  \delta = \frac{
    \sum_{i: \textrm{all D sites}} |n_i - 0.5|
  }{
    N/2
  },
\]
where $N$ is the number of the deuterium sites and $n_i$ is the occupancy of the $i$-th site (Please note that this index is nothing to do with thermodynamic parameters such as residual entropy). 
Fully disordered and ordered structures give $\delta=0$ and $1$, respectively.  

\reffig{fig:ice_alpha_ordering_Rietveld} summarises the outputs of Rietveld fitting.
The partially ordered $P2_12_12_1$ model resulted in the best fit whilst characteristic peaks were not reproduced well in the other models. 
The best $P2_12_12_1$ model achieved only partial ordering as $\delta = 0.12$ indicated, where only some particular deuterium sites had occupancies of 0 or 1. Slight orthorhombic lattice distortion was also suggested where $a=19.6972(23)$ \AA\ and $b = 19.6481(21)$ \AA. 
Refining the atomic coordinates first (prior to the occupancy refinement) did not improve the fitting.
Based on these results, we conclude that ice XXIII is a partially hydrogen-ordered phase with a reduced symmetry of $P2_12_12_1$\footnote{
   The hydrogen-ordered nature of ice XXIII was also supported by x-ray diffraction data below the phase transition temperature. X-ray diffraction profiles were nicely fitted with the oxygen-sublattice structure refined for ice XXI (\textit{e.g.}\ see \textit{ex-situ} x-ray data at 90 K in Extended Data Figure 3a). This indicates that oxygen-sublattice structure in ice XXIII almost retains that of ice XXI. 
}.

Since we separately refined the atomic coordinates and deuterium-site occupancies, we did not report errors for the refined parameters in the finalised model in Extended Table 3. 

\begin{oframed}
  \noindent
  \textit{Notes about occupancy refinement in GSAS}: We solved the equations of the deuterium-site occupancies arising from the Bernal–Fowler ice rules and refined occupancies under those constraints.
  We used a larger number of ordering parameters than the minimal independent variables required to describe the relations., \textit{i.e.}\ the definition of the ordering parameters was redundant. This was because we had to to reduce the number of the total crystallographic sites, due to a specification of the \textit{GSAS} program.
  For example, let us assume that there are four crystallographic sites (D0, D1, D2, D3) and their occupancies are described as $n_0=x,n_1=y,n_2=z,n_3=2-x-y-z$, respectively, whose initial values are all 1/2. This can be implemented in \textit{GSAS} by (i) splitting the D3 site into three sites (D3\_0, D3\_1, D3\_2), whose occupancies are $n_{30},n_{31},n_{32}$, respectively, and (ii) configuring the occupancy constraints of $\Delta n_0 = - \Delta n_{30} (= x), \Delta n_1 = - \Delta n_{31} (=y),  \Delta n_2 = - \Delta n_{32} (=z)$. To generalise this, when an occupancy is described by $n$ ordering parameters, the number of crystallographic sites needed to implement the occupancy constraints is also $n$. However, the maximum number of crystallographic sites allowed in \textit{GSAS} is 500. In low-symmetry models such as $P\bar{4}$, the number of crystallographic sites exceeds 500 when the minimal numbers of ordering parameters are used. This called for simplification of each occupancy representation --- rather than the use of the minimal number of ordering parameters ---, leading us to use a redundant expression of the site occupancies. Then we implemented additional conditions about ordering parameters by using the `Chemical Restraint' function (\textit{e.g.}\ the previous example can be rewritten as $n_0=x,n_1=y,n_2=z,n_3=w$, with an additional restraint equation of $x+y+z+w = 2$; the additional ordering parameter $w$ reduces the total number of sites from 6 to 4, with the addition of the restraint). It is noted that the maximum number of the `Chemical Restraints' in \textit{GSAS} is 9, which limits the number of redundant ordering parameters to be installed. 

  For Rietveld refinement in $P2_12_12_1$, we used 42 ordering parameters with 4 chemical restraints (\textit{i.e.}\ the minimal number of the required ordering parameter is 38, but we introduced 4 additional parameters). The number of total crystallographic sites was 462 ($<$500). 
  We had to allow slight violation of the ice rules, which is unavoidable when occupancies are refined under soft restraints. 
  D-site occupancies above 0.95 and below 0.05 were fixed to 1 and 0, respectively. 
  In the refined structural model, there are 303.8998 deuterium atoms and 152 oxygen atoms per unit cell, which give a chemical formula of D${}_{1.9994}$O.
  
\end{oframed}

A major weakness of our structural analysis was the failure to optimise the atomic coordinates \textit{simultaneously} and the use of (very heavy) restraints. 
Rietveld fitting is highly likely to be improved by refining the atomic coordinates simultaneously with the deuterium site occupancies in the low-symmetry models and/or by loosening the restraints (we are aware that some peak intensities are not perfectly reproduced by our model). 
However, this is `the better' --- and rather good --- structural model that can be analysed, as the structure is very complex and more detailed analyses are beyond the limits of powder diffraction.
We had to submit to such unusual stepwise refinement, but also argue that our $P2_12_12_1$ model captures major origins for the diffraction intensity changes (both atomic coordinates and deuterium-site occupancies) and thus essential structural features of the low-temperature phase ice XXIII.
If single crystal diffraction measurements of ice XXIII are possible in some way, this problem is likely to be overcome.

\clearpage

\subsection{Ice XXII}
\subsubsection{Unit cell and oxygen sublattice structure}
For ice XXII, similar stepwise analyses were performed. Autoindexing attempts by \textit{CONOGRAPH} for an x-ray diffraction data \fbox{Exp306-007} resulted in a face-centred orthorhombic lattice that gives $M_{30} > 20.7$, which is, in general, good enough to be regarded as the correct solution.
Space group candidates were constrained by checking the reflection conditions, and we checked all four leading candidates suggested by \textit{CONOGRAPH}: $Fddd$ [\#70], $Fdd2$ [\#43 setting 1], $Fd2d$ [\#43 setting 2], and $F2dd$ [\#43 setting 3].
We used the Pawley method to extract Bragg-reflection intensities using \textit{Z-Rietveld} software for experimental data \fbox{Exp306-007} and \fbox{Exp408: Sample02\_pos03-1}, followed by MEP optimisation and charge-flipping calculations performed similarly to ice XXI.
We tested all the space groups listed above, but in any case, charge-flipping calculations suggested $Fdd2$ [\#43, setting 1] structures regardless of the input space group symmetry and $Z$ number.

In $Fdd2$, the multiplicity of the highest-symmetry crystallographic position is 8 (Wyckoff position: 8a).
Thus, the number of atoms (and therefore the $Z$ number) is constrained to be a multiple of 8.
We tested several possible $Z$ numbers that give close density to ices V, VI, and VII, and found that the simulated powder x-ray diffraction profiles (using the charge-flipping outputs) agree well with experimental data only when $Z= 304$.
To double-check results, some of the ``wrong'' $Z$-number candidates were also examined in more detail. After Rietveld refinement starting from ``wrong $Z$-number'' structures, there were deficit or excess electron densities seen in the difference Fourier maps, suggesting that $Z$= 304 models are better than any other tested candidates. Thus, we concluded that ice XXII had an $Fdd2$ structure with $Z= 304$\footnote{
  $Z=304$ yields a density (at 250 K, 1.65 GPa) between ices VI and VII (\textit{i.e.}\ density: ice V < VI < XXII < <XXI < VII). 
}.

The best output from \textit{Superflip} was served for Rietveld refinement against an \textit{in-situ} x-ray diffraction profile \fbox{Exp306-007}.
Additionally, this oxygen-sublattice structure was confirmed by Rietveld analyses performed for the \textit{ex-situ} data collected with a recovered sample (Extended Data Figure 3b).
Only small movement of the oxygen atoms were confirmed before and after the Rietveld fitting, suggesting the charge-flipping output structure is nearly correct.

\subsubsection{H-bonded network from MD simulations}
Starting from this model, we run MD simulations to determine hydrogen-bonded network structure in ice XXII. 
The initial structure was determined based on Rietveld analyses for x-ray diffraction data (\fbox{EXP306-007}).
A short-range cut-off of 0.7 nm was applied for MD simulations of the unit cell containing 304 water molecules.
Flat-bottomed position restraints were imposed on the oxygen atoms, with a free radius of 0.05 nm and a harmonic force constant of 5000 kJ/mol/nm$^2$. 
Each water molecule was initially assigned a random orientation, and an NVT-MD simulation was performed at 250 K until the potential energy was stabilised. Subsequently, an NVT-MD simulation without the position restraints was conducted to obtain the average hydrogen positions.

All molecules formed tetrahedral hydrogen-bonded network structure and no complicated disorder --- as we encountered in ice XXI --- was observed in the time-averaged trajectory coordinates. 
A crystalloraphic structure model under the $Fdd2$ space-group symmetry was constructed. 
Hydrogen-atom positions suggested by MD simulations were compatible with the $Fdd2$ symmetry. 

\subsubsection{Rietveld analysis for neutron diffraction data --- finilising the $Fdd2$ model}
The final structural model was obtained by Rietveld fitting to a neutron diffraction profile.
 The only neutron diffraction experiment we have performed for ice XXIII is \fbox{Exp220}, where deuterated emulsified water (matrix:\ce{D2O} = 10:2 volume ratio) was isothermally compressed at 250 K up to 1.74 GPa (\fbox{Exp220\_HPN088031}), and thereafter quasi-isobarically cooled at 0.5 K/min to 90 K (\fbox{Exp220\_HPN088033}). 
 The 90 K data was slightly better than the 250 K data because of sharper peak widths, and better signal-to-noise ratio (due to a longer acquisition time; 24 hours at 250 K, 34 hours at 90 K). 
We checked the temperature dependence of the lattice parameters and found no abrupt changes during cooling, indicating the absence of temperature-induced first-order phase transitions (Extended Data Figure 7). 
Thus, the 90 K data \fbox{Exp220\_HPN088033} was used for Rietveld analysis. 

Full hydrogen disorder was assumed. The lattice parameters, atomic coordinates, atomic displacement parameters (common $U_\textrm{iso}$ values for the same element sites), profile parameters, and scaling factors were refined. Covalent-bonded O--D distances (0.955 $\pm$ 0.03 \AA) and some intermolecular and intramolecular D--D distances (with a moderate tolerance of $\pm$ 0.25 \AA) are restrained using `Restraints' fcuntion of GSAS with `Restraint Weight' of 300. 
Reasonably good fitting was achieved as reported in the main text\footnote{
  We found systematic deviation of the calculated diffraction intensities from observed one for lead. This is probably due to the presence of large grains and/or preferred orientation effects. The use of Le Bail method for lead would improve the fitting but we did not choose to do that. 
}.
We were not able to check potential hydrogen order due to structural complexity and bad signal-to-noise ratio of the neutron data (due to a reduced amount of aquaous phase compared to ice XXI data) but we suggest our Rietveld fit is good enough for a structural solution of such a complex unknown phase. 

\section{Supplementary methods: additional MD simulations}
\subsection{Melting curve of ice XXII from MD simulations (Extended Data Figure 5)}

The equilibrium melting temperature of ice XXII at 1.65 GPa was determined through the direct coexistence method \cite{Garcia_Fernandez2006-ns}.
The number of water molecules was 7296.
In the initial configuration, the solid region consisted of 2$\times$2$\times$3 of the unit cell of ice XXII (3648 molecules) and was combined with the liquid box with the same number of molecules.
The (001) plane was oriented perpendicular to the longest cell dimension.

The phase boundary between liquid water and ice XXII at lower pressures was calculated as follows as an extension of the calculation at 1.65 GPa.
We applied the Gibbs--Duhem scheme \cite{kofke1993gibbs} using temperature as an independent variable and calculating the ratio $\Delta h/\Delta v$ from independent simulations of bulk ice XXII and liquid water (2432 molecules for each phase), where $\Delta v$ and $\Delta h$ are the changes in molar volume and molar enthalpy during melting.
We used the classical fourth-order Runge--Kutta integration method with a step length of 10 K.
$\Delta v$ and $\Delta h$ were obtained from each 0.5 ns production NPT-MD simulation after 0.5 ns equilibration. 
RCFs were calculated from each 0.9 ns NPT-MD simulation after 100 ps equilibration, using 2432 molecules.

For reference, the phase boundary between liquid water and ice VI was determined by similar methods. The initial structure of ice VI was generated using the GenIce program \cite{Matsumoto2018-dh}. 
The system consisted of 50\% solid and 50\% liquid phases, with 3840 water molecules in total. 
The initial cell dimension was approximately 1:1:5.5. For the Gibbs-Duhem scheme, the ice VI and liquid phases were composed of 1250 and 2432 molecules, respectively.

\newpage
\section{Supplementary discussion}

\subsection{Solidification of the organic matrix phase}
Since emulsions are mixtures of oil, water, and surfactants, careful discussions are needed to interpret experimental results. We chose the 1:1 mixture of methylcyclohexane (MCH) and methylcyclopentane (MCP) as a matrix phase because (i) there are extensive experiments conducted with this matrix composition by Mishima and co-workers (\textit{e.g.}\ Refs.\ \cite{Mishima1996,Mishima1998,Mishima1998b,Mishima2001,Mishima2004,Mishima2011}) and (ii) Hauptmann \textit{et al.}\ (2016) reported on the basis of DSC measurements down to 93 K that the 1:1 MCP + MCH mixture did not undergo calorimetric vitrification and/or crystallisation at 1 atm \cite{Hauptmann2016}.

\begin{figure}
  \hrulefill
  \includegraphics{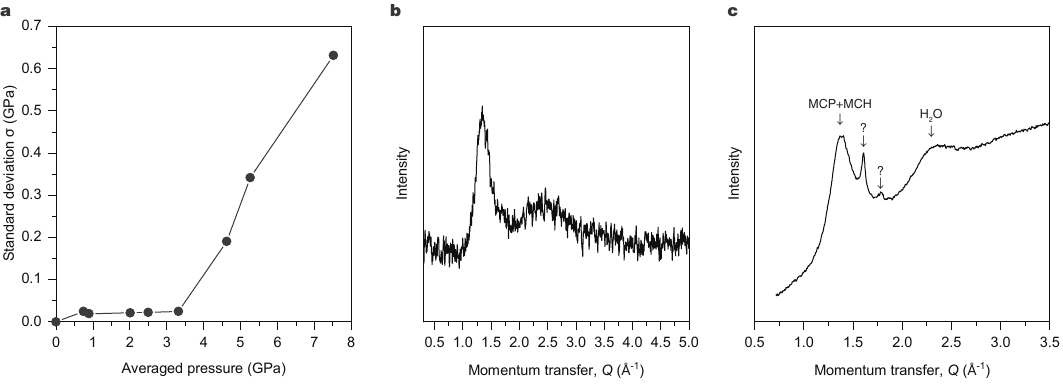}

  \caption{\textbf{Solidification of the organic matrix phase at room temperature.}
    \textbf{a}, Pressure dependence of the standard deviation $\sigma$ of pressure values estimated from 7 ruby spheres.
    \textbf{b}, X-ray diffraction pattern of the solidified 1:1 MCP + MCH mixture at the highest pressure (7.5 $\pm$ 0.6 GPa). Data was collected using a laboratory-source diffractometer (Rigaku R-AXIS IV++, Mo $K\alpha$ radiation) with a diamond anvil cell. The background intensity was collected with an empty cell and subtracted.
    \textbf{c}, Synchrotron x-ray diffraction patterns of an emulsified water sample around 2 GPa and room temperature. Additional broad features marked with `?' appear in emulsified water samples before the ice phase crystallises. Since these features never appear when the matrix phase is compressed without surfactant or when the amount of surfactant is reduced, we suggest that this originates from a combined nanostructure of the matrix and surfactant.}
  \label{fig:matrix_solidification}
\end{figure}

We conducted compression experiments at room temperature for the organic matrix components to check their pressure-induced solidification behaviour. Pressure-induced solidification was tracked by a useful method established by Klotz \textit{et al.}\ (e.g. Refs.\ \cite{Klotz2009,Klotz2009a}), where several ruby spheres are loaded in a diamond anvil cell, pressure values were calculated at each point using the ruby-fluorescence method \cite{Mao1978}, and their standard deviation $\sigma$ was monitored. Significant deterioration in $\sigma$ is observed upon solidification. In our experiment, 7 ruby spheres were loaded in a diamond anvil cell with the 1:1 MCP + MCH mixture. 
At 297 K, the solidification pressures of 1:1 mixture of MCP and MCH were determined to be 3.2 GPa, see \reffig{fig:matrix_solidification}a. 
This is sufficiently higher than the crystallisation pressure of the new ice polymorphs. 
Furthermore, even after the solidification of the 1:1 mixture of MCP and MCH, no sharp Bragg reflections were observed in the powder x-ray diffraction pattern shown in \reffig{fig:matrix_solidification}b, indicating its amorphous nature. 
Therefore, the matrix phase has little effect on our conclusions about the discovery and structural characterisation of the new phases of ice.

It is noted that in some experiments for W/O emulsions, we found broad features in powder x-ray diffraction patterns prior to the crystallisation of ice phases (indicated by '?' in \reffig{fig:matrix_solidification}c). 
These are surfactant-related because they were absent when pure MCP + MCH mixture was compressed. 
We confirmed that such broad features are not related to ice nucleation since sharp Bragg peaks of ice appeared at completely different pressures. 
We were not aware of this issue in the early stages of the project, and noticed that these additional unassigned peaks did not appear when the amount of surfactant was reduced from 3.5 wt\% to 2.0 wt\% of the matrix phase. 
We did not use peak intensities in the region where such a feature was observed for both phase recovery by the charge flipping method and the final structure refinement using neutron diffraction data, and therefore expect little effects of this feature on our structure analysis of new ice phases.

\subsection{Effects of the composition of emulsion samples on the crystallisation of unknown phases}
\begin{figure}
  \includegraphics{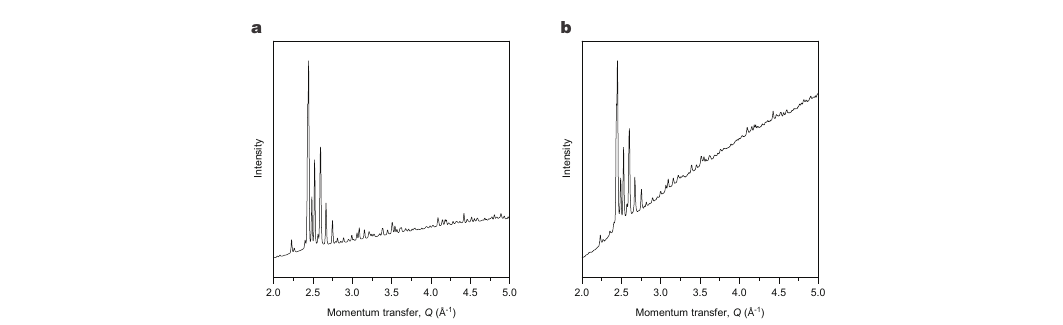}
  \caption{
    Testing effects of the matrix phase and surfactant on the crystallisation of ice XXI at room temperature. Ice XXI was crystallised at room temperature using different matrix/surfactant compositions and identified by x-ray diffraction measurements at BL-18C.
    \textbf{a}, Matrix: 1:1 methylcyclohexane + methylcyclopentane mixture; Surfactant: sorbitan tristearate.
    \textbf{b}, Matrix: ethylcyclohexane; Surfactant: DIS-F.
  }
  \label{fig:compare_composition}
\end{figure}

We also investigated the effect of surfactant and organic matrix molecules on crystallisation. 
If the molecules in emulsions other than water play critical roles in the nucleation of ice, changing the composition of the emulsion samples should lead to different pressure-induced crystallisation products. 
Although MCP and MCH are insoluble in water, such effects must be considered carefully to validate physical interpretations about the phase transition behaviour of water. 
We conducted some reference experiments for the pressure-induced crystallisation of ice XXI at room temperature:

\begin{itemize}
  \item Emulsified water was prepared with a different matrix (ethylcyclohexane: ECH) and a different surfactant (DIS-F, kindly provided by Sakamoto Yakuhin Kogyo Co., Ltd.) and experiments of compression-induced crystallisation were performed. 
  However, all the emulsion samples with different chemical compositions crystallised to the same phase, ice XXI, when compressed at room temperature (\reffig{fig:compare_composition}), indicating that the surfactant and matrix molecules do not affect the crystallisation of ice by, for example, acting as a nucleator of ice XXI. 
  Powder x-ray diffraction patterns of the ECH/DIS-F emulsion and the MCP+MCH/sorbitan tristearate emulsions indicate the same crystallisation product: ice XXI. 
  Although the molecular structures of MCP, MCH, and ECH are similar, the powder diffraction patterns cannot be perfectly the same if these organic molecules are incorporated in the crystals we report as an ice polymorph.
  \item The intensity of the broad diffuse scattering feature of the organic matrix did not drastically change when sharp Bragg peaks appeared. This suggests that the organic matrix is not incorporated in the crystals (if matrix-water compounds, namely MCP-MCH clathrate hydrates, are formed, MCP/MCH molecules with a comparable molar amount should be ``consumed'' for crystallisation, causing changes in the diffuse scattering intensities).
  \item In addition, it is noted that at lower pressures, emulsified water crystallises into established ice phases such as ices III and IV. 
  Mishima and co-workers have extensively studied the melting curves of all the ice phases that melts into liquid water below 1.5 GPa by using emulsified water with the same composition \cite{Mishima1984a,Mishima2011}. This supports that emulsified water can be used to probe structural changes of bulk water phases.
\end{itemize}

\subsection{Ice XXI and Lee \textit{et al.}'s `unknown phase'}

In this section, we note that ice XXI may be obtained in non-emulsified water. 

Recently, Lee \textit{et al.}\ posted a preprint article on \textit{Research Square} \cite{Lee2024a}. They performed extensive compression experiments of water at room temperature using piezo-actuated dynamic diamond anvil cells (dDACs), reporting that there are five types of pressure-induced crystallisation pathways, which occur in stochastic manners. One of them involves the crystallisation of an `unknown phase'. They identified this phase by time-resolved powder x-ray diffraction measurements at the HED instrument of European XFEL. 

\begin{figure*}
  \includegraphics{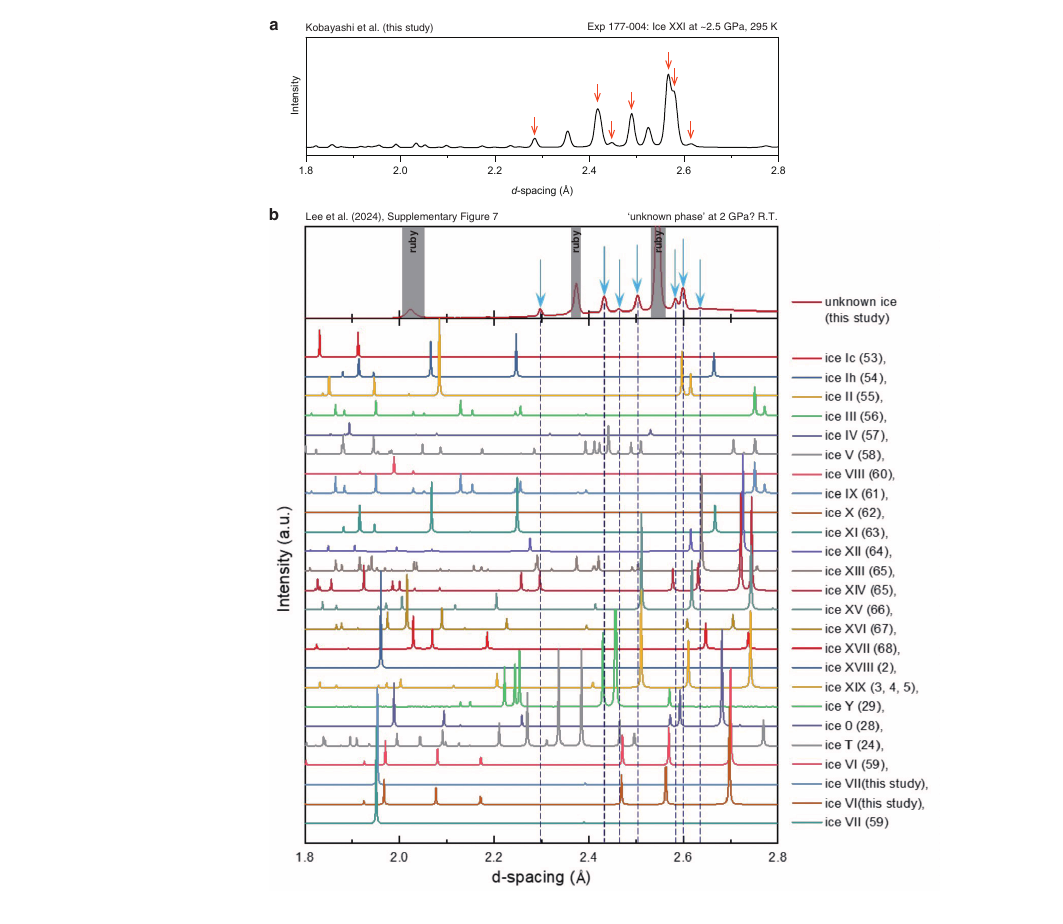}
  \caption{
    \textbf{a}, Our x-ray diffraction pattern of ice XXI.
    \textbf{b}, X-ray diffraction patterns of the `unknown phase' reported by Lee \textit{et al.}, plotted with simulated patterns of some known and predicted phases. Adopted from Supplementary Figure 7 of Ref.\ \cite{Lee2024a} under the CC BY 4.0 licence.
  }  
  \label{fig:Lee_et_al}
\end{figure*}

X-ray diffraction patterns of their `unknown phase' collected at European XFEL are somewhat similar to ours of ice XXI (\reffig{fig:Lee_et_al}). However, detailed comparison between our and their diffraction data is extremely challenging due to heavy overlap of the `unknown phase' peaks with ruby and ice VII peaks. 
Intensities of the most intense features around  $2.5\; \textrm{\AA}<d<2.6\; \textrm{\AA}$ do not match well between the data of the two groups. 
This may be simply attributed to the differences in the crystal structure originating from the pressure difference; the `unknown phase' of Lee \textit{et al.}\ crystallised at \textasciitilde2 GPa, which is considerably lower  than our ice XXI in emulsified water. 
Detailed comparison and powder pattern fitting are desired using our model.

\newpage
\section{Supplementary notes: ice nomenclature for the new phases discovered in this study}
Nowadays, it is generally accepted to assign a new Roman numeral to a phase of ice when it has been experimentally established. A criterion for the Roman numeral assignement commonly accepted in the community was given by Petrenco and Whitworth, who stated `\textit{We take the view that the series of Roman numerals should only be used for crystalline phases that are experimentally well established, for example by crystallography or spectroscopy}' (Ref.\ \cite{Physics_of_Ice}, Chapter 11). 
As we reported the structure models by crystallographic methods, there should be no doubt that the new ice polymorphs we discovered fulfil the requirements for assigning Roman numerals. However, we are aware of an ongoing controversy about which Roman numeral is the latest for the accepted ice polymorphs. The latest phase of ice should be either

\begin{itemize}
  \item 	Ice XIX (nineteen): a low-temperature counterpart of ice VI forms at high pressures, identified by powder neutron diffraction to be a distinct phase because (i) ice XIX has the $\sqrt{2} \times \sqrt{2} \times 1$  supercell of ice VI and (ii) the space group of ice XIX differs from neither ice VI nor another low-temperature counterpart of ice VI (ice XV), by Gasser \textit{et al.}\ (2021) \cite{Gasser2021a}, Yamane \textit{et al.}\ (2021) \cite{Yamane2021}, and Salzmann \textit{et al.}\ (2021) \cite{Salzmann2021}. Although the structure of ice XIX remains controversial yet (partial hydrogen order or full disorder), it is clear that ice XIX is a thermodynamically distinct phase of ice because diffraction data clearly evidence that ice XIX has a different crystallographic symmetry from the related phases, ices VI and XV.

  \item Ice XX (twenty): a superionic phase of ice having the body-centred cubic (bcc) oxygen sublattice, identified with the Roman numeral by x-ray diffraction by Prakapenka \textit{et al.}\ (2021) \cite{Prakapenka2021}. It is noted that the bcc-superionic phase was reported by other groups, \textit{e.g.}\ Queyroux \textit{et al.}\ (2020) \cite{Queyroux2020}, in a non-Roman-numbered form. It has the same oxygen sublattice structure as ice VII, but has a different density (equation of state), making it ice VII and superionic bcc ice (ice XX) distinguishable by x-ray diffraction, according to Prakapenka \textit{et al.}\ (2021). In addition, optical conductivity data by Prakapenka \textit{et al.}\ may support the superionicity.
\end{itemize}

An issue to be discussed lies with `ice XX'. Experimental data for `ice XX' by Prakapenka \textit{et al.}\ \cite{Prakapenka2021} at least supports that the bcc phase might be superionic, \textit{i.e.}\ hydrogen atoms are mobile while oxygen atoms maintain a periodic lattice. Nevertheless, we are aware of an opinion in the ice community in which the experimental evidence presented in the paper by Prakapenka \textit{et al.}\ (and other preceding/succeeding works) is not sufficient to distinguish ``normal'' bcc ice (ice VII) from the potentially superionic high-temperature phase(s). In particular, we raise concerns regarding the relation between what is called `ice XX' and what is called `plastic ice VII' in the recent report by Rescigno \textit{et al.}\ (2025) \cite{Rescigno2025}. At present, it is unclear whether `ice XX' differs from `plastic ice VII' since both are ``dynamic bcc ice phases'', which have been suggested to be different from ``normal'' ice VII. In this context, `being different from ice VII' (in density, for example) is insufficient. Therefore, at least (i) direct evidence for superionicity (\textit{i.e.}\ proton dynamics in ice) and (ii) discontinuous behaviour of thermodynamic parameters that supports first-/second-order phase transitions (between ``normal'' and superionic bcc ice, and between superionic and plastic bcc ice) is needed for the firm identification of ice XX. We agree that Prakapenka \textit{et al.}`s 'ice XX' is highly likely to be a thermodynamically distinct phase of ice displaying dynamic disorder, but we also agree with the opinion that more direct evidence for superionicity is desired. 
If we assign the Roman numeral XX to our new phase in the present work, scientific confusion with two `ice XX' is expected in the future ice community. Such confusion should be avoided, as the ice community experienced similar situations regarding multiple `ice XI' \cite{Benoit1996,Kawada1972} and `ice XII' \cite{sirota1994phase,Svishchev1996,Lobban1998a} in the last century (\textit{also see} Chapter 11 of Ref.\ \cite{Physics_of_Ice}).

Therefore, we herein propose to
\begin{enumerate}
  \item 	Assign the Roman numeral XXI, instead of XX, to our new phase that crystallises from supercooled water at room temperature and 2.4 GPa, and by heating emulsified HDA at 2.5 GPa.
  \item 	Treat the Roman numeral XX as a reserved number for the ``dynamically disordered'' bcc phase at 20–30 GPa. Whether different Roman numerals are to be assigned to the plastic and superionic ices should be decided based on the pressure dependence of the proton dynamics in the bcc ice phase(s) at high temperatures. If there is a clear phase boundary between the superionic and plastic phases, they should be named differently (the superionic side should be called ice XX).
\end{enumerate}

\newpage
\printbibliography

\end{document}